\documentclass[12pt]{iopart}

\usepackage{graphicx,color,amssymb,bm,dcolumn}
\usepackage{epsfig}
\usepackage{bm}
\usepackage{dsfont}
\newcommand{\bra}[1]{\langle#1|}
\newcommand{\ket}[1]{|#1\rangle}

\def\bbm[#1]{\mbox{\boldmath $#1$}}

\usepackage[english]{babel}

\def\bbm[#1]{\mbox{\boldmath $#1$}}
\newcommand{\TE}{\mathrm{TE}}
\newcommand{\TM}{\mathrm{TM}}

\begin{document}

\title[Creation and protection of entanglement  in  systems out of thermal equilibrium]{Creation and protection of entanglement  in  systems out of thermal equilibrium}

\author{Bruno Bellomo$^{1,2}$  and Mauro Antezza$^{1,2,3}$}

\address{$^1$Universit\'{e} Montpellier 2, Laboratoire Charles Coulomb UMR 5221 - F-34095, Montpellier, France}
\address{$^2$CNRS, Laboratoire Charles Coulomb UMR 5221 - F-34095, Montpellier, France}
\address{$^3$Institut Universitaire de France - 103, bd Saint-Michel
F-75005 Paris, France}
\ead{bruno.bellomo@univ-montp2.fr}
\ead{mauro.antezza@univ-montp2.fr}

\begin{abstract}
We investigate the creation of entanglement between two quantum emitters interacting with a realistic  common stationary electromagnetic field out of thermal equilibrium.
In the case of two qubits we show that the absence of equilibrium allows the generation of steady entangled states, which is  inaccessible at thermal equilibrium and is realized without any further external action on the two qubits. We first give a simple physical interpretation of the phenomenon in a specific case and then we report a detailed investigation on the dependence of the entanglement dynamics on the various physical parameters involved. Sub- and super-radiant effects are discussed, and qualitative differences in the dynamics concerning both creation and protection of entanglement according to the initial two-qubit state are pointed out.

\end{abstract}
\maketitle


\section{Introduction}

Quantum systems may present correlations of both quantum and classical nature. Entanglement captures quantum correlations due to the non separability of the system state \cite{Werner89,Amico,Horodecki09}. The presence of these correlations is connected to the rise of non local effects in quantum theory \cite{Einstein35} and has been recognized as a key resource in several fields of quantum technology, including quantum computing \cite{BookNielsen}, quantum cryptography \cite{Acin}, quantum teleportation \cite{Bennett} and quantum metrology \cite{Giovannetti}. A main obstacle to the concrete exploitation of quantum features in the above applications is the detrimental effects of environmental noise \cite{BookBreuer}. The unavoidable coupling with degrees of freedom of the surrounding environment generally leads to a decay of quantum coherence properties \cite{Zurek03}, preventing the possible exploitation of quantum correlations present in the system.

A considerable effort has been done to understand the effects of environmental noise on the dynamics of correlations present in an open quantum system \cite{Diosi, Yu04, BellomoPRL07, BellomoPRA10,BellomoPRA12}, and to contrast the natural fragility of quantum coherence properties \cite{Lidar98, Bellomo08,ManiscalcoPRL08}. Reservoir engineering methods have pointed out the possibility to change the perspective from reducing  the coupling with the environment to modifying the environmental properties in order to manipulate the system of interest thanks to its proper dissipative dynamics
\cite{Plenio, Cirac09, Brune11}. Other approaches exploit the effect of measurements and feedback to drive the systems towards a target state \cite{Mancin07,Carvalho11}.

A possible way to create quantum correlations between two systems is to make them interact with a common environment \cite{Braun}, which can also cause a
revival of entanglement  \cite{Tanas08}. In the case of two emitters  in a common vacuum or thermal electromagnetic field, in absence of matter close to them, the mediated interaction plays a role over distances of the order of the common transition wavelength \cite{Agarwal1974, FicekBook2005}. It has been evidenced that the presence of plasmonic waveguides near the emitters can allow a mediated interaction over larger distances \cite{Dzsotjan} whose effect on the entanglement dynamics has been discussed  \cite{Vidal2011}. However, at thermal equilibrium the dynamical creation of entanglement eventually ceases at some time and the system thermalizes towards a thermal state which is a classical mixture. Steady entanglement can be instead generated by adding the action of an external driving laser \cite{Ficeck2011}.

The influence of several independent reservoirs at different temperatures, whose emission does not depend on their internal structure (material or geometry), has been considered in several contexts, including generation of entanglement
in nonequilibrium steady states,  both in the case of few spins \cite{Quiroga, Huang09, Camalet2011} and of a chain of spins \cite{ Znidaric, Manzano, Camalet2013} and in the context of quantum thermal machines \cite{Linden10, Linden11}.

However, in a realistic configuration the actual reflection and transmission properties of the bodies surrounding the quantum emitters should be taken into account, and may become particularly relevant if the emitters are placed close to the bodies (near-field effects).
New possibilities emerging in such realistic systems out of thermal equilibrium have been recently pointed out in different contexts ranging from heat transfer \cite{Joulain, MesAnt12}, to Casimir-Lifshits forces \cite{AntezzaPRL05, AntezzaJPA06, AntezzaPRL07, AntezzaPRA08, Bimonte2009, KardarPRL2011}.
There,  radiation fields out of thermal equilibrium in configurations of quite general nature have been characterized in terms of  the correlators of the total field depending on the scattering matrices of the bodies composing the total system \cite{MesAntEPL11, MesAntPRA11}. In the case of single emitters in such environments, new tools exploiting the absence of thermal equilibrium to manipulate the atomic dynamics realizing inversion of population and cooling of internal atomic temperature have been pointed out \cite{BellomoEPL2012,BellomoPRA2013}. Recently,  the  case of two quantum emitters has also been analyzed, pointing out a new remarkable mechanism to generate and protect entanglement in a steady way in systems out of thermal equilibrium \cite{Bellomo2013}.

 In this paper, we report a detailed investigation of this phenomenon by studying the internal dynamics of a system composed by two quantum emitters (real atoms or artificial ones as quantum dots or superconducting qudits) placed in front of an arbitrary body embedded in a thermal radiation whose temperature is different from that of the body. The paper is organized as follows. In Sec. \ref{par:model} we describe the physical model under investigation and we derive a master equation for the general case of  two $N$-level emitters. In Sec. \ref{par:one body} we derive closed-form expressions  for the functions governing the dynamics, in terms of the scattering matrices of the body and valid for {\it arbitrary geometrical and material properties}.
In Sec. \ref{par: atom in front of a slab} we develop these expressions in
 the case when the body is a slab of finite thickness.
From Sec. \ref{par:twoqubits} on we specialize our analysis to the case of a two-qubit system, comparing  cases in and out of thermal equilibrium. We point out the occurrence of peculiar phenomena emerging out of thermal equilibrium such as the generation of steady entanglement and a simple interpretation for this phenomenon is presented for a particularly interesting case.  The general case of arbitrary values of the parameters is then  discussed in Sec. \ref{par:numerical investigation}.  In Sec. \ref{par:Conclusions} we draw our conclusions.

\section{Model}\label{par:model}

\begin{figure}[b]
\begin{center}\includegraphics[width=0.55\textwidth]{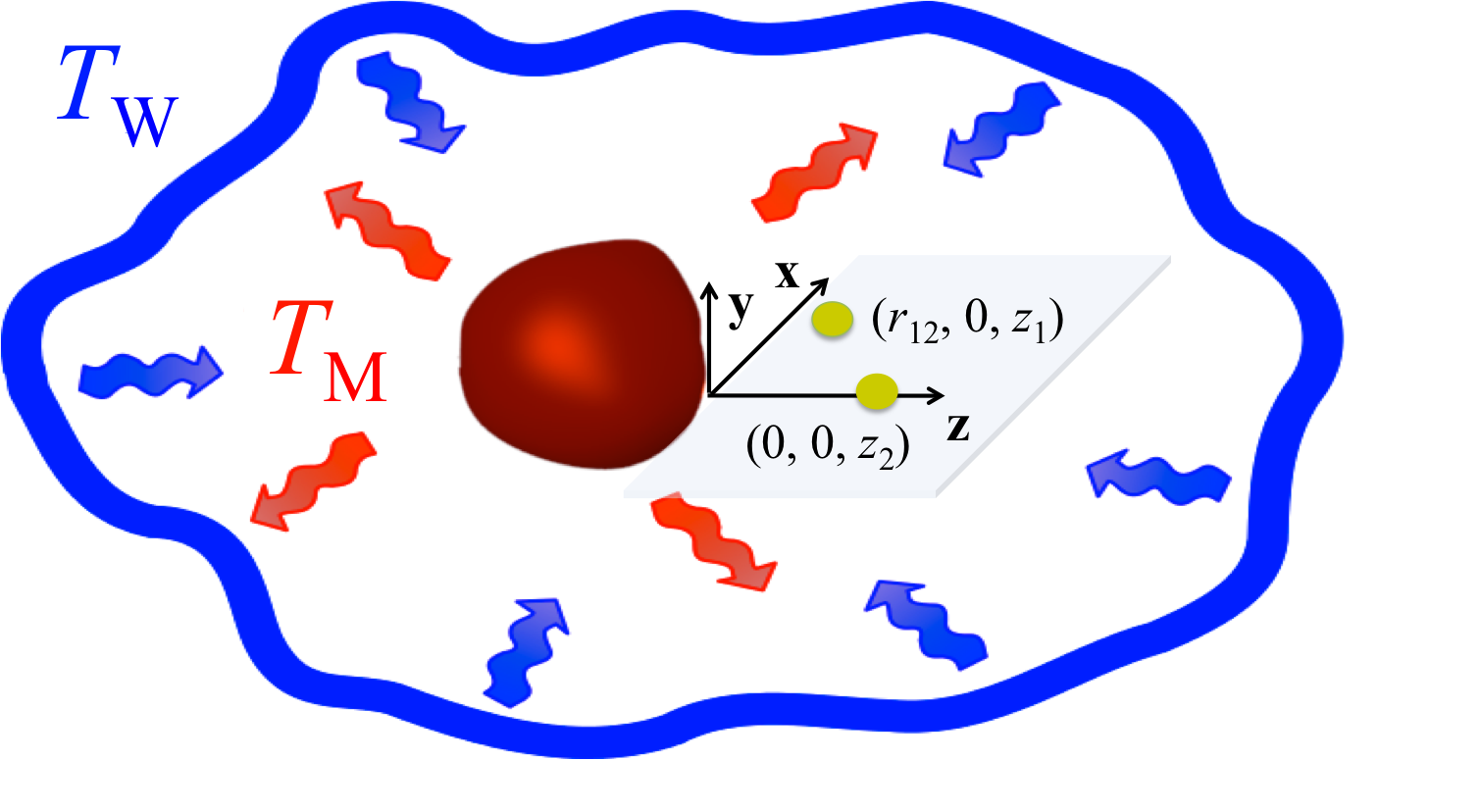}\end{center}
\caption{\label{fig:1}\footnotesize  Physical configuration: two quantum emitters close to an arbitrary body whose temperature $T_\mathrm{M}$ is kept fixed and different from that of the surrounding walls $T_\mathrm{W}$. The two emitters are placed in $\mathbf{R}_1=(\mathbf{r}_1, z_1)$ and $\mathbf{R}_2=(\mathbf{r}_2, z_2)$, where $\mathbf{r}_1$ and $\mathbf{r}_2$ are vectors in the $xy$ plane. In the figure, we choose the $x$ axis along the direction $\mathbf{r}_1-\mathbf{r}_2$ and $x_2=0$, naming $r_{12}=|\mathbf{r}_1-\mathbf{r}_2|=x_1$ (this choice of the reference system is used in Sec. \ref{par: atom in front of a slab} in the specific case when the body is a slab).}
\end{figure}

We consider a system made of two quantum emitters $q=1, 2$ interacting with an environment consisting of an electromagnetic field  which is stationary and out of thermal equilibrium. This is generated by the field emitted by a body (M) at temperature $T_\mathrm{M}$ of arbitrary geometry and dielectric response  and by the field emitted by the far surrounding walls (W) at temperature $T_\mathrm{W}$, which is eventually transmitted and reflected by the body itself (see figure \ref{fig:1}).
$T_\mathrm{M}$ and $T_\mathrm{W}$ are kept fixed in time, realizing a stationary configuration for the electromagnetic field. The surrounding walls have an irregular shape and are distant  enough from the body and the emitters so that their field can be treated at the emitters' locations, in absence of the body, as a blackbody radiation independent from their composition. This is not true for the field emitted by the body \textrm{M} which cannot be treated as a blackbody since its radiation depend on its actual properties as its geometry and its dielectric function. The total Hamiltonian has the form
\begin{equation}\label{Hamiltonian}H=H_S+H_E+H_I,\end{equation}
where $H_S$ and $H_E$ are the free Hamiltonians of the two emitters and of the environment.
The interaction between the emitters and the field, in the multipolar coupling and in dipole approximation, is \cite{CohenTannoudji97}
\begin{equation}\label{interaction Hamiltonian}H_I=-\sum_q \mathbf{D}_q\cdot\mathbf{E}(\mathbf{R}_q) ,\end{equation}
where $\mathbf{D}_q$ is the electric-dipole operator of emitter $q$ and $\mathbf{E}(\mathbf{R}_q) $ is the electric field at its position $\mathbf{R}_q$.

We first consider the general case in which each emitter has $N_q$ internal levels $\ket{n}_q$ where $n \in \{1, ... , N_q\}$ of frequency $\omega_{n}^q$ (ordered by increasing energy).  Given  two arbitrary levels $n$ and $m$, their frequency difference  is indicated by $\omega_{nm}^q=\omega_{n}^q-\omega_{m}^q$ and  the transition matrix element of the dipole operator by $\mathbf{d}_{mn}^q={}_q\bra{m}\mathbf{D}_q\ket{n}_q$. The free Hamiltonian of the two emitters is
\begin{equation}\label{Atom hamiltonian N}H_S=\sum_{q=1}^2 H_q=\sum_{q, \epsilon_n^q} \epsilon_n^q \Pi(\epsilon_n^q),\end{equation}
where $\Pi(\epsilon_n^q)=\ket{n}_q {}_q\bra{n}$ are the projectors associated to each eigenvalue $\epsilon_n^q=\hbar\omega_{n}^q$ (possibly degenerate) of $H_q$.
The dipole operator of emitter $q$ in the interaction picture, $\mathbf{D}_q(t)=\exp(\frac{i}{\hbar}H_St)\mathbf{D}_q\exp(-\frac{i}{\hbar}H_St)$, results to be
\begin{equation}\label{dipole operator N}\mathbf{D}_q(t)=\sum_{^{\,m,n}_{n>m}}\Bigl(\mathbf{d}_{m n}^q\ \sigma_{mn}^q e^{-i\omega_{nm}^qt}+\mathrm{h.c.}\Bigr),\end{equation}
where $\sigma_{mn}^q=\ket{m}_q{}_q\bra{n}$ and $\omega_{n m}^q \ge 0$.
By moving to the interaction picture, we obtain for $H_I$
\begin{equation}\label{Interaction hamiltonian}H_I(t)=-\sum_q \mathbf{D}_q(t)\cdot\mathbf{E}(\mathbf{R}_q,t),\end{equation}
where the time-dependent electric field is given by   $\mathbf{E}(\mathbf{R}_q,t)=\exp(\frac{i}{\hbar}H_Et)E(\mathbf{R}_q)\exp(-\frac{i}{\hbar}H_Et)$. In the following each mode of the field is identified by the frequency $\omega$, the transverse wave vector $\mathbf{k}=(k_x,k_y)$, the polarization index $p$ (taking the values $p=1,2$ corresponding to transverse-electric
(TE), and transverse-magnetic (TM) polarizations respectively), and the direction or propagation $\phi=\pm1$ (shorthand notation $\phi=\pm$) along the $z$ axis [see figure \ref{fig:1}]. In this approach, the total wavevector takes the form $\mathbf{K}^\phi=(\mathbf{k},\phi k_z)$, where the $z$ component of the wavevector $k_z$ is a dependent variable given by $k_z=\sqrt{\frac{\omega^2}{c^2}-k^2}$, where $k=|\mathbf{k}|$. The explicit expression of the field at an arbitrary point $\mathbf{R}$ is
\begin{equation}\label{electric field}\mathbf{E}(\mathbf{R},t)=2\mathrm{Re}\Biggl[\int_0^{+\infty}\frac{d\omega}{2\pi}e^{-i\omega t}\mathbf{E}(\mathbf{R},\omega)\Biggr],\end{equation}
where a single-frequency component reads
\begin{equation}\label{Eomega}\mathbf{E}(\mathbf{R},\omega)=\sum_{\phi,p}\int\frac{d^2\mathbf{k}}{(2\pi)^2}e^{i\mathbf{K}^\phi\cdot\mathbf{R}}\hat{\bbm[\epsilon]}^\phi_{p}(\mathbf{k},\omega)E^\phi_p(\mathbf{k},\omega),\end{equation}
where $E^\phi_p(\mathbf{k},\omega)$ is the field amplitude operator associated to the mode $(\omega, \mathbf{k},p,\phi)$.
For the TE and TM polarization vectors appearing in \eref{Eomega} we adopt the following standard definitions
\begin{equation}\eqalign{
\hat{\bbm[\epsilon]}^\phi_\TE(\mathbf{k},\omega)&=\hat{\mathbf{z}}\times\hat{\mathbf{k}}=\frac{1}{k}(-k_y\hat{\mathbf{x}}+k_x\hat{\mathbf{y}}),\\
\hat{\bbm[\epsilon]}^\phi_\TM(\mathbf{k},\omega)&=\frac{c}{\omega}\hat{\bbm[\epsilon]}^\phi_\TE(\mathbf{k},\omega)\times\mathbf{K}^{\phi}=\frac{c}{\omega}(-k\hat{\mathbf{z}}+\phi k_z\hat{\mathbf{k}}),}\end{equation}
where $\hat{\mathbf{x}}$, $\hat{\mathbf{y}}$ and $\hat{\mathbf{z}}$ are the unit vectors along the three axes and $\hat{\mathbf{k}}=\mathbf{k}/k$.

\subsection{Master equation}\label{par:derivation}

The starting point to study the dynamics of the two emitters is, in the interaction picture, the von Neumann equation for the total density matrix $\rho_{\mathrm{tot}}(t)$:
\begin{equation}\label{von Neummann equation}\frac{d}{dt}\rho_{\mathrm{tot}}(t)=-\frac{i}{\hbar}[H_I(t),\rho_{\mathrm{tot}}(t)].\end{equation}
The reduced density matrix of the two emitters is given by $\rho =\mathrm{Tr}_E[\rho_{\mathrm{tot}}]$, where $\mathrm{Tr}_E$ denotes the trace over the degrees of freedom of the environment. To derive a master equation for $\rho$ we follow the procedure described in \cite{BookBreuer} for the case of one emitter by extending it to our system made of two emitters.
We name $\omega_q$ an arbitrary  transition frequency of emitter $q$ (positive and negative). In general, several transitions can be characterized by the same frequency $\omega_q$ both because of degeneracy and/or the occurrence of equidistant levels.
We rewrite each cartesian component  of the dipole operator, $[\mathbf{D}_q]_i$ ($i=\{x, y, z\}$), as
\begin{equation}\label{definitionA}
[\mathbf{D}_q]_i= \sum_{\omega_q}  \sum_{^{\:\:\:\:\{\epsilon^{q}_{n},\epsilon_m^q\}}_{\epsilon^{q}_{n}-\epsilon_m^q=\hbar \omega_q}}
\Pi(\epsilon_m^q)[\mathbf{D}_q]_i\Pi(\epsilon^{q}_{n})=\sum_{\omega_q} A^q_i(\omega_q),
\end{equation}
where $A^q_{i}(\omega_q)$ and $A^{q\,\dag}_{i}(\omega_q)$
turn out to be eigenoperators of $H_q$ with frequencies $- \omega_q$ and $+\omega_q$, respectively, i.e. $[H_q,A_i^q(\omega_q)]=-\omega_q A_i^{q}(\omega_q)$ and $[H_q,A_i^{q\,\dag}(\omega_q)]=+\omega_q A_i^{q\,\dag}(\omega_q)$. It also holds  $A_i^{q\,\dag}(\omega_q)=A_i^{q}(-\omega_q)$ and  $\exp(\frac{i}{\hbar}H_St) A^q_i(\omega_q)\exp(-\frac{i}{\hbar}H_St)=e^{-i\omega_q t}A^q_{i}(\omega_q) $. In the central term of  \eref{definitionA}, the first sum  is over all the frequencies $\omega_q$ while the second is over all the couples of energy eigenvalues $\epsilon^{q}_{n}$ and $\epsilon_m^q$ of $H_q$ such that  $\epsilon^{q}_{n}-\epsilon_m^q=\hbar \omega_q$.
Following  \cite{BookBreuer}, it is useful to rewrite
 $H_I(t)$ of  \eref{Interaction hamiltonian} in terms of the eigenoperators $A^q_{i}(\omega_q)$ as
\begin{equation}\label{Interaction hamiltonian 2}H_I(t)=-\sum_q\sum_{i,\omega_q}e^{-i\omega_q t}A^q_{i}(\omega_q) E_i(\mathbf{R}_q,t).\end{equation}
From  \eref{definitionA} it follows that the vector $\mathbf{A}^q(\omega_q)=\{A^q_{x}(\omega_q), A^q_{y}(\omega_q), A^q_{z}(\omega_q)\}$ is given by
\begin{equation}\label{A general form}
\mathbf{A}^q(\omega_q)=  \sum_{^{\:\:\:\{m,n\}}_{\omega_{nm}^q=\omega_q}}
 \mathbf{d}_{mn}^q \sigma_{mn}^q=\mathbf{A}^{q\,\dag}(-\omega_q),
\end{equation}
where
the sum is over all the couples $n$ and $m$ such that $\omega_{nm}^q=\omega_q$.
By applying to the case of two emitters the standard procedure for the microscopic derivation of a master equation reported  in \cite{BookBreuer}, under Born, Markovian and rotating-wave approximations \footnote{The Born-Markov approximation is
typically valid in the weak coupling regime when the bath correlation time is small compared to the
relaxation time of the system. Under rotating wave approximation  rapidly oscillating terms
 can be neglected when  the inverse of frequency
differences involved in the problem are small compared to the
relaxation time of the system (see appendix A of \cite{BellomoPRA2013} for a more detailed discussion).},  one can obtain (using also the condition $\langle E_i(\mathbf{R},t)\rangle=0$) in the Schr\"{o}dinger representation:
\begin{equation}\label{master equation}   \eqalign{&\frac{d}{d t}\rho =-\frac{i}{\hbar} [H_S,\rho ]-i
\sum_{q, q', \omega}\sum_{i,i'}  \Big\{s_{ii'}^{q q'}(\omega) [A^{q\,\dag}_{i}(\omega)A_{i'}^{q'}(\omega),\rho ]   \\
&\,+\gamma_{ii'}^{qq'}(\omega) \Big(A^{q'}_{i'}(\omega)\rho A^{q\,\dag}_{i}(\omega) -\frac{1}{2}\{A^{q\,\dag}_{i}(\omega)A^{q'}_{i'}(\omega),\rho \}\Big)\Big\},}\end{equation}
where $\omega \gtreqless 0$, being terms with positive or negative $\omega$ associated, respectively, to downward and upward transitions.
In the above equation,  for $q\neq q'$ the sum $\sum_{q, q', \omega}$ is over all common frequencies $\omega_q=\omega_{q'}=\omega$  (this condition derives from the rotating wave approximation) while  for $q= q'$ it is over all transition frequencies of each emitter, and $\gamma_{ii'}^{qq'}(\omega)$, and $s_{ii'}^{q q'}(\omega)$ are defined by
\begin{equation}\label{Xi function}\eqalign{
\gamma_{ii'}^{qq'}(\omega) &=\Xi_{ii'}^{qq'}(\omega) +\Xi_{i' i }^{q' q \,*}(\omega)  , \quad s_{ii'}^{qq'}(\omega) =  \frac{\Xi_{ii'}^{qq'}(\omega) -\Xi_{i' i }^{q' q \,*}(\omega)}{2 i}, \\
\Xi_{ii'}^{qq'}(\omega)&= \frac{1}{\hbar^2}\int_0^\infty  \!\!ds\,e^{i\omega s}\langle E_i(\mathbf{R}_q,t)E_{i'}(\mathbf{R}_{q'},t-s)\rangle ,}\end{equation}
where the field correlation functions enter in the function $\Xi_{ii'}^{qq'}(\omega)$.
It follows that $\Xi_{ii'}^{qq'}(\omega)=\frac{1}{2}\gamma_{ii'}^{qq'}(\omega)+i s_{ii'}^{qq'}(\omega)$,  $[\gamma_{ii'}^{qq'}(\omega)]^*=\gamma_{i'i}^{q'q}(\omega)  $ and $[s_{ii'}^{qq'}(\omega)]^*=s_{i'i}^{q'q}(\omega)$.

The initial state of the total system in  \eref{master equation} is assumed to be factorized, $\rho_{\mathrm{tot}}(0)=\rho(0)\rho_E$. In the case $\rho_E$ is a stationary state of the environment ($[H_E,\rho_E]=0$) the correlation functions are homogenous in time, that is $\langle E_i(\mathbf{R}_q,t)E_{i'}(\mathbf{R}_{q'},t-s)\rangle=\langle E_i(\mathbf{R}_q,s)E_{i'}(\mathbf{R}_{q'},0)\rangle$, so that
\begin{equation}\label{gammaijdef}\gamma_{ii'}^{qq'}(\omega)=\frac{1}{\hbar^2}\int_{-\infty}^\infty ds\,e^{i\omega s}\langle E_i(\mathbf{R}_q,s)E_{i'}(\mathbf{R}_{q'},0)\rangle\end{equation}
does not depend on time. The functions defined in  \eref{Xi function}  appearing in the master equation \eref{master equation} depend thus only on the field correlation functions $\langle E_i(\mathbf{R}_q,s)E_{i'}(\mathbf{R}_{q'},0)\rangle$, whose computation out of thermal equilibrium will be the subject of Secs. \ref{par:one body} and \ref{par: atom in front of a slab}.

We now explicitly write the master equation  \eref{master equation} in the case  of absence of degenerate and equidistant levels in each emitter, when
the definition of eigenoperators $\mathbf{A}^q(\omega_q)$  \eref{A general form} reduces to  $\mathbf{A}^q(\omega_q)= \mathbf{d}^q_{mn} \sigma_{mn}^q$ (to each $\omega_q$ corresponds only one couple of energy eigenvalues $\{ \epsilon^{q}_{m}, \epsilon_n^q\}$).
To this purpose, we develop  the sum  over $\omega$ in  \eref{master equation}, which for each $|\omega|$ runs over $\omega$ (downward transitions) and $-\omega$ (upward transitions), as $\sum_\omega f(\omega)=\sum_{\omega>0}  f(\omega) + \sum_{\omega>0}  f(-\omega)$. From now on $\omega$ indicates always a positive frequency and we drop ``$>0$'' in the sums over $\omega$. Introducing this new convention and
using the explicit form for $A^{q}_{i}(\omega)$  \eref{A general form}, we can recast   \eref{master equation}  as
\begin{equation}\label{master equation 5}\eqalign{&\frac{d}{d t}\rho =-\frac{i}{\hbar}[H_S,\rho ]-i
\sum_{q, \omega} \Bigl\{S^{qq}(\omega)[\sigma_{nn}^q,\rho ]+S^{qq}(-\omega)[\sigma_{mm}^q,\rho ] \Bigr\}\\&\,-i\sum_{q\neq q', \omega}\Lambda^{qq'}(\omega)[\sigma_{mn}^{q\,\dag}\sigma_{m'n'}^{q'},\rho ]
+\sum_{q, q', \omega}\Bigl\{\Gamma^{qq'}(\omega)\Big(\sigma^{q'}_{m'n'}\rho \sigma_{mn}^{q\,\dag}\\&\,-\frac{1}{2}\{\sigma^{q\,\dag}_{mn}\sigma^{q'}_{m'n'},\rho \}\Big)+\Gamma^{qq'}(-\omega)\Big(\sigma^{q'\dag}_{m'n'}\rho \sigma^{q}_{mn}-\frac{1}{2}\{\sigma^{q}_{mn} \sigma^{q'\,\dag}_{m'n'},\rho \}\Big)\Bigr\} ,}\end{equation}
where the sum $\sum_{q, q', \omega}$ in the second line is relative to all transition frequencies of each emitter for $q=q'$ and only  to the common transition frequencies for $q\neq q'$,  $(m,n)$ and $(m',n')$ individuate respectively the transition of each emitter corresponding to the frequency $\omega$, and  we have defined the functions
\begin{equation}\label{me parameters N}\eqalign{
\fl S^{qq}(\omega)&=\sum_{i,i'} s_{ii'}^{qq}(\omega)[\textbf{d}_{mn}^q]^*_{i} [\textbf{d}_{mn}^{q}]_{i'},\quad
S^{qq}(-\omega)=\sum_{i,i'}s_{ii'}^{qq}(-\omega)[\textbf{d}_{mn}^q]_{i}[\textbf{d}_{mn}^q]^*_{i'},
\\
\fl \Lambda^{qq'}(\omega)&= \sum_{i,i'}  [\mathbf{d}_{mn}^q]_i^* [\mathbf{d}_{m'n'}^{q'}]_{i'}
 \bigl[s_{ii'}^{q q'}(\omega)+s_{i'i}^{q' q}(-\omega)\bigr],
\\
\fl \Gamma^{qq'}(\omega)&=\sum_{i,i'}\gamma_{i i'}^{qq'}(\omega)[\textbf{d}_{mn}^q]^*_{i} [\textbf{d}_{m'n'}^{q'}]_{i'},\quad
\Gamma^{qq'}(-\omega)=\sum_{i,i'}\gamma_{i i'}^{qq'}(-\omega)[\textbf{d}_{mn}^q]_{i}[\textbf{d}_{m'n'}^{q'}]^*_{i'}.}\end{equation}
 We remark that it holds  $[\Lambda^{qq'}(\omega)]^*=\Lambda^{q'q}(\omega)$ and $[\Gamma^{qq'}(\omega)]^*=\Gamma^{q'q}(\omega)$.
In   \eref{master equation 5}, function $\Lambda^{qq'}(\omega)$ represents a coherent (dipole-dipole) interaction between
the emitters mediated by the field while dissipative effects enter through the $\Gamma$ functions.
In particular, $\Gamma^{qq'}(\pm\omega)$ are individual ($q=q'$) and common field-mediated collective ($q\neq q'$) emitter transition rates, related to both quantum and thermal fluctuations of the electromagnetic field at the emitters' position.

\section{Emitters close to an arbitrary body}\label{par:one body}

Here we derive the field correlation functions needed to compute the functions in \eref{me parameters N} for non equilibrium configurations in the case of an arbitrary body and multilevel emitters. These functions will depend on the two temperatures $T_\mathrm{M}$ and $T_\mathrm{W}$ and on the material and geometrical properties of the body as well. We follow the derivation discussed in \cite{MesAntPRA11} in the more general case of two bodies and three temperatures and the derivation relative to a single quantum emitter in the presence of a single body and two temperatures \cite{BellomoPRA2013}. Here we extend the latter derivation to the case of two quantum emitters. Some of the computations involved are reported in \ref{app:one body}.

\begin{figure}[t!]
\begin{center}\includegraphics[width=0.55\textwidth]{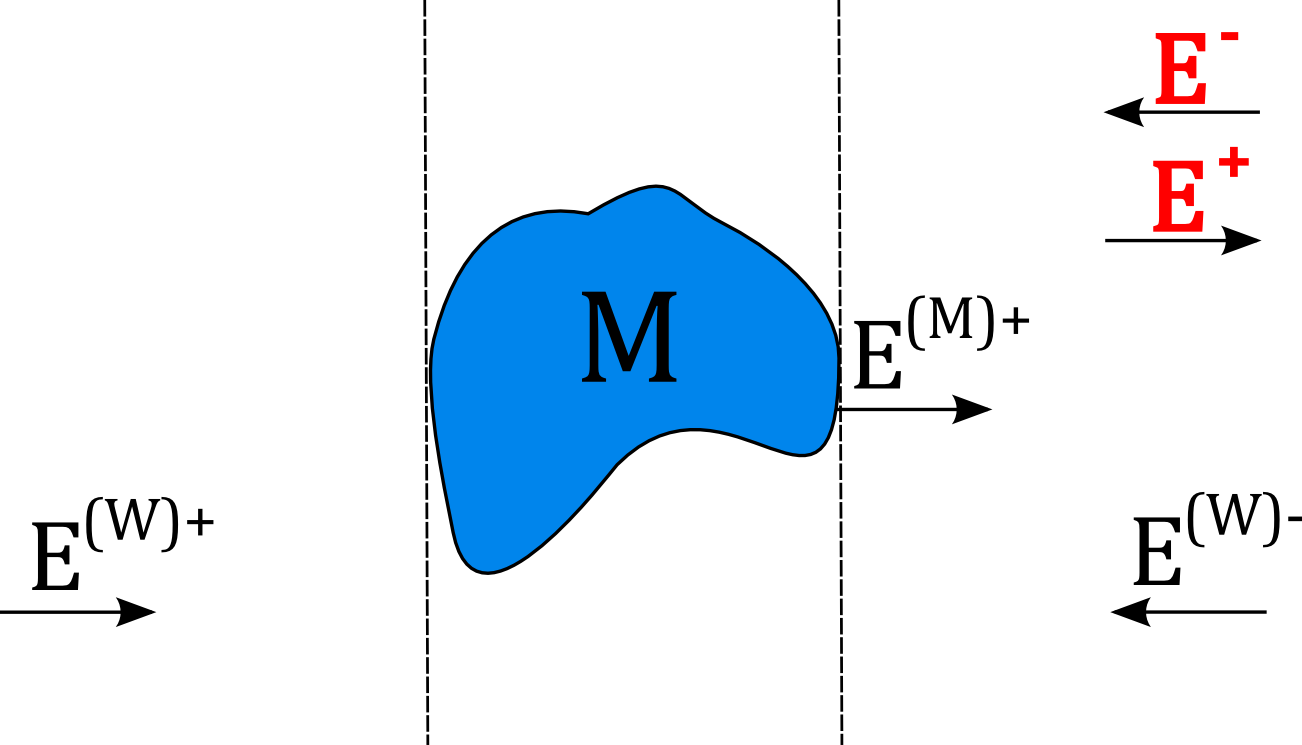}\end{center}
\caption{\label{fig:varicampi}\footnotesize  $E^{\pm}$ are the total field in the zone on the right of the body $\mathrm{M}$. $E^{(\mathrm{M})+}$ is the field emitted by the body towards the right, while $E^{(\mathrm{W})+}$ and $E^{(\mathrm{W})-}$ are  the fields emitted by the surrounding walls (not shown in the picture) coming  respectively from the left and from the right, eventually impinging on the body.}
\end{figure}

The starting point is to decompose, on the right side of the body where the emitters are located, the  amplitude operators of total field modes  propagating in the two directions $ z>0$ and $z<0$ in terms of the fields emitted by the surrounding walls (W) and by the body (M).
For a given set $(\omega,\mathbf{k},p)$, we have for the two directions
\begin{equation}\label{totalfield}
E^+=E^{(\mathrm{M})+}+\mathcal{T}E^{\mathrm{(W)}+}+\mathcal{R}E^{\mathrm{\mathrm{(W)}}-}, \quad E^-=E^{\mathrm{(W)}-},
\end{equation}
where we made the dependence on $\omega$, $\mathbf{k}$ and $p$ implicit.
The total field $E^-$ propagating toward the body (i.e. toward the left) is equal to the field  emitted by the walls $E^{(\mathrm{W})-}$ coming from the left, while the total field $E^+$ propagating toward the right  results from the field $E^{(\mathrm{M})+}$ directly produced by the body, the transmission through the body of the field $E^{(\mathrm{W})+}$ emitted by the walls coming from the left, and the reflection by the body of the field $E^{(\mathrm{W})-}$ coming from the right [see figure \ref{fig:varicampi}].
The operators $\mathcal{R}$ and $\mathcal{T}$ are the reflection and transmission scattering operators associated to the right side of the body, whose explicit definition can be found for example in \cite{MesAntPRA11}. They connect any outgoing (reflected or transmitted) mode of the field to the entire set of incoming modes. By using  \eref{totalfield} one can write the total field correlators in terms of the correlators of the fields emitted by each source.

The source fields have been characterized as in \cite{MesAntPRA11}  by assuming  that for the body M and the walls W a local temperature which  remains constant in time can be defined and that the emission process of the body  is essentially not influenced by the presence of the external radiation impinging on the body itself. This assumption leads to the hypothesis that the part of the total field emitted by the body is the same as it would be if the body were at thermal equilibrium with the environment at its own temperature so that the correlators of the field emitted by each body can still be deduced using the fluctuation-dissipation theorem at its local temperature.

Under this assumption, the following symmetrized correlation functions [$\langle AB\rangle_\mathrm{sym}=(\langle AB\rangle+\langle BA\rangle)/2$] have been derived
\begin{equation}\label{sourcefields}\eqalign{ \fl &\langle E^{\mathrm{(M)}+}_p(\mathbf{k},\omega)E^{\mathrm{(M)}+\dag}_{p'}(\mathbf{k}',\omega')\rangle_\mathrm{sym}=\frac{\omega}{2\epsilon_0c^2}N(\omega,T_\mathrm{M})
2\pi\delta(\omega-\omega')\bra{p,\mathbf{k}}\Bigl(\mathcal{P}_{-1}^\mathrm{(pw)}\\
\fl &\, -\mathcal{R}\mathcal{P}_{-1}^\mathrm{(pw)}\mathcal{R}^{\dag} +\mathcal{R}\mathcal{P}_{-1}^\mathrm{(ew)} -\mathcal{P}_{-1}^\mathrm{(ew)}\mathcal{R}^{\dag}-\mathcal{T}\mathcal{P}_{-1}^\mathrm{(pw)}\mathcal{T}^{\dag}\Bigr)\ket{p',\mathbf{k}'}, \\
\fl & \langle E^{\mathrm{(W)}\phi}_p(\mathbf{k},\omega)E^{\mathrm{(W)}\phi'\dag}_{p'}(\mathbf{k}',\omega')\rangle_\mathrm{sym}
=\frac{\omega}{2\epsilon_0c^2}N(\omega,T_\mathrm{W}) 2\pi\delta(\omega-\omega')\delta_{\phi,\phi'}\bra{p,\mathbf{k}}\mathcal{P}_{-1}^{\mathrm{(pw)}}\ket{p',\mathbf{k}'},}\end{equation}
where we have introduced
\begin{equation}\label{NadnP}\eqalign{
\fl &N(\omega,T)=\frac{\hbar\omega}{2}\coth\Bigl(\frac{\hbar\omega}{2k_BT}\Bigr)=\hbar\omega\Bigl[\frac{1}{2}+n(\omega,T)\Bigr],\quad n(\omega,T)=\Bigl(e^{\frac{\hbar\omega}{k_BT}}-1\Bigr)^{-1},\\
\fl &\bra{p,\mathbf{k}}\mathcal{P}_n^\mathrm{(pw/ew)}\ket{p',\mathbf{k}'}=k_z^n\bra{p,\mathbf{k}}\Pi^\mathrm{(pw/ew)}\ket{p',\mathbf{k}'}.}\end{equation}
In the above equation $\Pi^\mathrm{(pw)}$ and $\Pi^\mathrm{(ew)}$ are the projectors on the propagative ($c\,k<\omega$, corresponding to a real $k_z$) and evanescent ($c\,k>\omega$, corresponding to a purely imaginary $k_z$) sectors respectively. By combining  \eref{totalfield} and \eref{sourcefields}, in  \ref{app:one body} a general expression for the total correlation functions in frequency space has been derived in  \eref{funzcorr}.
This expression can be used to compute the functions $ \gamma_{ii'}^{qq'}(\omega),  \gamma_{ii'}^{qq'}(-\omega)$  and $s_{ii'}^{qq'}(\omega)$ entering in \eref{me parameters N}, by exploiting  their connection with the  correlation functions between frequency components of the total field given in  \eref{gammaij2}.

To move to the final expression of the functions in \eref{me parameters N} we first rewrite the antinormally ordered correlation functions  \eref{funzcorr} as
\begin{equation}\label{funzcorr2}\eqalign{
\langle E_i & (\mathbf{R}_q,\omega)  E_{i'}^\dag(\mathbf{R}_{q'},\omega)\rangle=\frac{\hbar \omega^3}{3 \pi\epsilon_0 c^3}\Big\{[1+n(\omega,T_\mathrm{W})]  \\&\times [\alpha_\mathrm{W}^{q q'}(\omega)]_{ii'}+ [1+n(\omega,T_\mathrm{M})][\alpha_\mathrm{M}^{q q'}(\omega)]_{ii'} \Big\},
}\end{equation}
from which the normally ordered correlation functions are obtained  by replacing $\bigl[1+n(\omega,T_i)\bigr]$ with $n(\omega,T_i)$ and by taking the complex conjugate (this procedure derives from Kubo's prescription as explained in \ref{app:one body})
\begin{equation}\label{funzcorr3}\eqalign{
\langle E_i^\dag&(\mathbf{R}_q,\omega)  E_{i'}(\mathbf{R}_{q'},\omega)\rangle=\frac{\hbar \omega^3}{3 \pi\epsilon_0 c^3}\Big\{n(\omega,T_\mathrm{W}) \\&
\times [\alpha_\mathrm{W}^{q q'}(\omega)]_{ii'}^*+ n(\omega,T_\mathrm{M})][\alpha_\mathrm{M}^{q q'}(\omega)]_{ii'}^*\Big\},
}\end{equation}
and where we have introduced two $\alpha$ functions which do not depend on temperatures and on dipoles, and depend on the geometrical and material properties of the body through the operators $\mathcal{R}$ and $\mathcal{T}$:
\begin{equation}\label{alphaW}\eqalign{\fl
[\alpha_\mathrm{W}^{q q'}(\omega)]_{ii'}&=\frac{3\pi c}{2 \omega}\sum_{p,p'}\int\frac{d^2\mathbf{k}}{(2\pi)^2}\int\frac{d^2\mathbf{k}'}{(2\pi)^2}e^{i(\mathbf{k}\cdot\mathbf{r}_q-\mathbf{k}'\cdot\mathbf{r}_{q'})} \bra{p,\mathbf{k}}\Bigl\{e^{i(k_z z_q-k_z^{'*}z_{q'})}
\\ \fl
& \, \times [\hat{\bbm[\epsilon]}_p^+(\mathbf{k},\omega)]_i[\hat{\bbm[\epsilon]}_{p'}^{+}(\mathbf{k}',\omega)]_{i'}^*\Bigl(\mathcal{T}\mathcal{P}_{-1}^{\mathrm{(pw)}}\mathcal{T}^{\dag}+\mathcal{R}\mathcal{P}_{-1}^{\mathrm{(pw)}}\mathcal{R}^{\dag}\Bigr)\\ \fl
& \,+e^{i(k_zz_q+k_z^{'*}z_{q'})}[\hat{\bbm[\epsilon]}_p^+(\mathbf{k},\omega)]_i[\hat{\bbm[\epsilon]}_{p'}^{-}(\mathbf{k}',\omega)]_{i'}^*\mathcal{R}\mathcal{P}_{-1}^{\mathrm{(pw)}}\\ \fl
& \,+e^{-i(k_zz_q+k_z^{'*}z_{q'})}[\hat{\bbm[\epsilon]}_p^-(\mathbf{k},\omega)]_i[\hat{\bbm[\epsilon]}_{p'}^{+}(\mathbf{k}',\omega)]_{i'}^*\mathcal{P}_{-1}^{\mathrm{(pw)}}\mathcal{R}^{\dag}\\ \fl
&+e^{-i(k_z z_{q}-k_z^{'*}z_{q'})}[\hat{\bbm[\epsilon]}_p^-(\mathbf{k},\omega)]_i[\hat{\bbm[\epsilon]}_{p'}^{-}(\mathbf{k}',\omega)]_{i'}^*\mathcal{P}_{-1}^{\mathrm{(pw)}}\Big\}\ket{p',\mathbf{k}'},\\ \fl
[\alpha_\mathrm{M}^{q q'}(\omega)]_{ii'}& =\frac{3\pi c}{2 \omega}\sum_{p,p'}\int\frac{d^2\mathbf{k}}{(2\pi)^2}\int\frac{d^2\mathbf{k}'}{(2\pi)^2}e^{i(\mathbf{k}\cdot\mathbf{r}_q-\mathbf{k}'\cdot\mathbf{r}_{q'})}\bra{p,\mathbf{k}}\Bigl\{e^{i(k_z z_q-k_z^{'*}z_{q'})}
\\ \fl
&\,\times [\hat{\bbm[\epsilon]}_p^+(\mathbf{k},\omega)]_i[\hat{\bbm[\epsilon]}_{p'}^{+}(\mathbf{k}',\omega)]_{i'}^* \Bigl[\Bigl(\mathcal{P}_{-1}^\mathrm{(pw)}-\mathcal{R}\mathcal{P}_{-1}^\mathrm{(pw)}\mathcal{R}^{\dag}+\mathcal{R}\mathcal{P}_{-1}^\mathrm{(ew)}\\ \fl
&\,-\mathcal{P}_{-1}^\mathrm{(ew)}\mathcal{R}^{\dag}-\mathcal{T}\mathcal{P}_{-1}^\mathrm{(pw)}\mathcal{T}^{\dag}\Bigr)\,\Big\}\ket{p',\mathbf{k}'}.}\end{equation}
Functions $[\alpha^{qq'}_\mathrm{W}(\omega)]_{ii'}$ and $[\alpha^{qq'}_\mathrm{M}(\omega)]_{ii'}$ are  {in general}  complex satisfying $\mathrm{Im} [\alpha^{qq'}_\mathrm{W}(\omega)]_{ii'}= -\mathrm{Im} [\alpha^{qq'}_\mathrm{M}(\omega)]_{ii'}$. The last property assures that  the function $[\alpha^{qq'}_\mathrm{W}(\omega)]_{ii'}+ [\alpha^{qq'}_\mathrm{M}(\omega)]_{ii'}$ is real as expected, being proportional to the imaginary part of the Green's function [see  \eref{IMofG}].

Now we can compute the transition rates in \eref{me parameters N}, using  \eref{gammaij2}, \eref{funzcorr2} and  \eref{funzcorr3},
\begin{equation}\label{gamma functions finale}\eqalign{
\fl  \Gamma^{qq'}(\omega)& = \sqrt{\Gamma_0^q(\omega)\Gamma_0^{q'}(\omega) } \Big\{[1+n(\omega,T_\mathrm{W})]\alpha_\mathrm{W}^{q q'}(\omega) + [1+n(\omega,T_\mathrm{M})]\alpha_\mathrm{M}^{q q'}(\omega)\Big\}\\
\fl   \Gamma^{qq'}(-\omega)& =  \sqrt{\Gamma_0^q(\omega)\Gamma_0^{q'}(\omega) } \Big[n(\omega,T_\mathrm{W})  \alpha_\mathrm{W}^{q q'}(\omega)^* +n(\omega,T_\mathrm{M})\alpha_\mathrm{M}^{q q'}(\omega)^*\Big]
 ,}\end{equation}
where $\Gamma_0^q(\omega)=\frac{|\textbf{d}_{mn}^q|^2\omega^3}{3  \hbar  \pi\epsilon_0 c^3} $ is the vacuum spontaneous-emission rate of transition $\ket{n}_q\to\ket{m}_q$ of emitter $q$ and we have introduced the new functions
\begin{equation}\label{alpha Me eW  finale}\eqalign{ \fl
\alpha_\mathrm{W}^{q q'}(\omega) =\sum_{i,i'}[\tilde{\textbf{d}}_{mn}^q]^*_{i} [\tilde{\textbf{d}}_{m'n'}^{q'}]_{i'} [\alpha_\mathrm{W}^{q q'}(\omega)]_{ii'} \quad
\alpha_\mathrm{M}^{q q'}(\omega) =\sum_{i,i'}[\tilde{\textbf{d}}_{mn}^q]^*_{i} [\tilde{\textbf{d}}_{m'n'}^{q'}]_{i'} [\alpha_\mathrm{M}^{q q'}(\omega)]_{ii'},
}\end{equation}
being $[\tilde{\textbf{d}}_{mn}^q]_{i}=[\textbf{d}_{mn}^q]_{i}/|\textbf{d}_{mn}^q|$. Differently from $[\alpha_\mathrm{W(M)}^{q q'}(\omega)]_{ii'}$, the functions $\alpha_\mathrm{W(M)}^{q q'}(\omega)$ depend on the choice of emitters' dipoles. In the case of two qubits, which will be treated in Secs. \ref{par:twoqubits} and \ref{par:numerical investigation}, there is only one transition for each emitter and above equations \eref{gamma functions finale} and \eref{alpha Me eW  finale}  hold with
the notation $\textbf{d}_{mn}^q=\textbf{d}^q$.

With regards to the function $\Lambda^{qq'}(\omega)$ we obtain,  using  \eref{gammaij2}, \eref{funzcorr2} and  \eref{funzcorr3},
\begin{equation}\label{lambda2}\eqalign{
\fl \Lambda^{qq'}(\omega)= \frac{\sqrt{\Gamma_0^q(\omega)\Gamma_0^{q'}(\omega) }}{\omega^3} \sum_{i,i'}  [\tilde{\mathbf{d}}_{mn}^q]_i^*[\tilde{\mathbf{d}}_{m'n'}^{q'}]_{i'}
 \mathcal{P}\int_{-\infty}^{+\infty}\frac{\omega'^3 d\omega'}{2\pi}\frac{[\alpha_\mathrm{W}^{q q'}(\omega')]_{ii'} +[\alpha_\mathrm{M}^{q q'}(\omega')]_{ii'} }{\omega-\omega'} ,
 }
 \end{equation}
 where we used the properties $[\alpha_\mathrm{W(M)}^{q q'}(\omega')]_{ii'} =[\alpha_\mathrm{W(M)}^{q' q}(\omega')]_{i'i}^*$ and $[\alpha_\mathrm{W(M)}^{q q'}(-\omega')]_{ii'}=[\alpha_\mathrm{W(M)}^{q q'}(\omega')]^*_{ii'}$.
It follows that $\Lambda^{qq'}(\omega)$ does not depend on the presence or absence of thermal equilibrium, being independent on the temperatures.
Using the relation between $\alpha$ functions of  \eref{alphaW} and the Green's function of the system in \eref{IMofG} derived in \ref{par:Green function}, the integration over frequencies in   \eref{lambda2} can be done by using the Kramers-Kronig relations connecting real and imaginary parts of the Green's function:
\begin{equation}\label{lambda3}\eqalign{ \fl
\qquad \:\Lambda^{qq'}(\omega) &= -\frac{1}{\hbar} \sum_{i,i'}  [\mathbf{d}_{mn}^q]_i^* [\mathbf{d}_{m'n'}^{q'}]_{i'}
 \mathcal{P}\int_{-\infty}^{+\infty} \frac{d\omega'}{\pi} \frac{\mathrm{Im} \,G_{ii'} (\mathbf{R}_q,\mathbf{R}_{q'},\omega')}{\omega'-\omega} \\
   \fl & = -\frac{1}{\hbar} \sum_{i,i'}  [\mathbf{d}_{mn}^q]_i^* [\mathbf{d}_{m'n'}^{q'}]_{i'}  \mathrm{Re} \, G_{ii'} (\mathbf{R}_q,\mathbf{R}_{q'},\omega).
   }
 \end{equation}

\section{Emitters close to a slab}\label{par: atom in front of a slab}

\begin{figure}[t]
\begin{center}\includegraphics[width=0.55\textwidth]{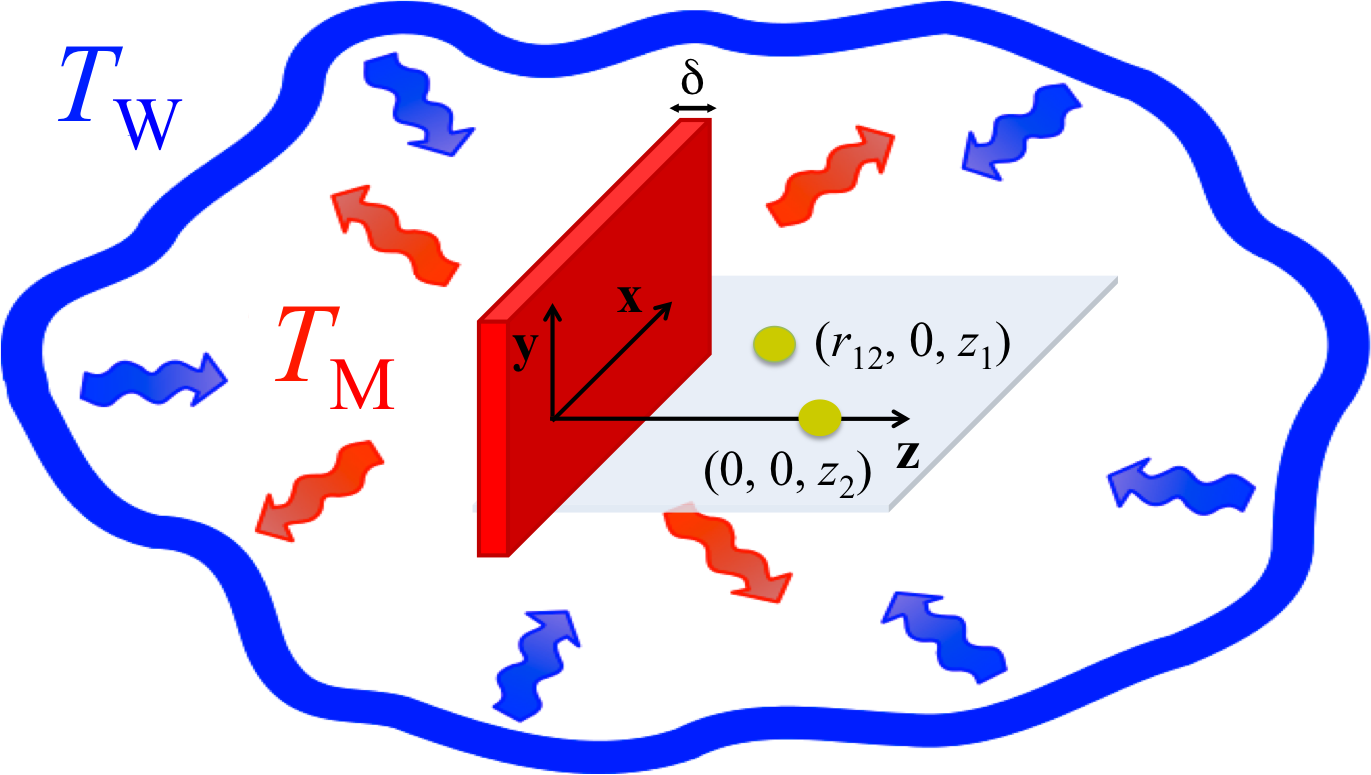}\end{center}
\caption{\label{fig:slab}\footnotesize  Two quantum emitters in front of a slab of thickness $\delta$ at a fixed temperature $T_\mathrm{M}$, surrounded by walls kept at a temperature $T_\mathrm{W}$.}
\end{figure}

We now specialize the derivation of previous section to the case when the body is a slab of finite thickness $\delta$, defined by the two interfaces $z=0$ and $z=-\delta$ (see figure \ref{fig:slab}).
In this simple case, explicit expressions for the transmission and reflection operators can be exploited \cite{MesAntEPL11,MesAntPRA11}.
Because of the translational invariance of a planar slab with respect to the $xy$ plane, the slab reflection and transmission operators, $\mathcal{R}$ and $\mathcal{T}$, are diagonal and equal to
\begin{equation}\label{RT1slab}\eqalign{\bra{p,\mathbf{k}}\mathcal{R}\ket{p',\mathbf{k}'}&=(2\pi)^2\delta(\mathbf{k}-\mathbf{k}')\delta_{pp'}\rho_{p}(\mathbf{k},\omega),\\
\bra{p,\mathbf{k}}\mathcal{T}\ket{p',\mathbf{k}'}&=(2\pi)^2\delta(\mathbf{k}-\mathbf{k}')\delta_{pp'}\tau_{p}(\mathbf{k},\omega),}\end{equation}
where the Fresnel reflection and transmission coefficients modified by the finite thickness $\delta$ are given by  (we recall that $p=1,2$ corresponding to TE and TM polarizations)
\begin{equation}\eqalign{\rho_{p}(\mathbf{k},\omega)&=r_{p}(\mathbf{k},\omega)\frac{1-e^{2ik_{zm}\delta}}{1-r_{p}^2(\mathbf{k},\omega)e^{2ik_{zm}\delta}},\\
\tau_{p}(\mathbf{k},\omega)&=\frac{t_{p}(\mathbf{k},\omega)\bar{t}_{p}(\mathbf{k},\omega)e^{i(k_{zm}-k_z)\delta}}{1-r_{p}^2(\mathbf{k},\omega)e^{2ik_{zm}\delta}}.\\}\end{equation}
In the previous equations we have introduced the $z$ component of the $\mathbf{K}$ vector inside the medium,
\begin{equation}k_{zm}=\sqrt{\varepsilon(\omega)\frac{\omega^2}{c^2}-\mathbf{k}^2},\end{equation}
$\varepsilon(\omega)$ being the dielectric permittivity of the  slab, the ordinary vacuum-medium Fresnel reflection coefficients
\begin{equation}r_{\mathrm{TE}}=\frac{k_z-k_{zm}}{k_z+k_{zm}},\qquad r_{\mathrm{TM}}=\frac{\varepsilon(\omega)k_z-k_{zm}}{\varepsilon(\omega)k_z+k_{zm}},\end{equation}
as well as both the vacuum-medium (noted with $t$) and medium-vacuum (noted with $\bar{t}$) transmission coefficients
\begin{equation}\eqalign{t_{\mathrm{TE}}&=\frac{2k_z}{k_z+k_{zm}},\qquad\hspace{.3cm}t_{\mathrm{TM}}=\frac{2\sqrt{\varepsilon(\omega)}k_{z}}{\varepsilon(\omega)k_z+k_{zm}},\\
\bar{t}_{\mathrm{TE}}&=\frac{2k_{zm}}{k_z+k_{zm}},\qquad\bar{t}_{\mathrm{TM}}=\frac{2\sqrt{\varepsilon(\omega)}k_{zm}}{\varepsilon(\omega)k_z+k_{zm}}.}\end{equation}

After replacing the matrix elements \eref{RT1slab}  in  \eref{alphaW}  we obtain for the $\alpha$ functions [we choose the $x$ axis along the vector $\mathbf{r}_1-\mathbf{r}_{2}$ whose coordinates in the plane $x y$ are then $(r_{12},0)$, being $r_{12}=|\mathbf{r}_1-\mathbf{r}_{2}|$],
\begin{equation}\label{alphaslab}\eqalign{\fl [\alpha_\mathrm{W}^{qq'}(\omega)]_{ii'} =& \frac{3c}{8\pi\omega}\sum_{p}\Biggl\{\int_0^{\frac{\omega}{c}}\frac{dk\,k}{k_z}
\Bigl[  e^{ik_z (z_q-z_{q'})}  [N_p^{qq'}(k,\omega)]^{++}_{ii'} \bigl(|\rho_{p}(\mathbf{k},\omega)|^2+|\tau_{p}(\mathbf{k},\omega)|^2\bigr)\\ \fl
&+e^{ik_z(z_q+z_{q'})} [N_p^{qq'}(k,\omega)]^{+-}_{ii'}\rho_{p}(\mathbf{k},\omega)+e^{-ik_z(z_q+z_{q'})}[N_p^{qq'}(k,\omega)]^{-+}_{ii'}\rho_{p}(\mathbf{k},\omega)^*\\ \fl
&+e^{-ik_z (z_q-z_{q'})} [N_p^{qq'}(k,\omega)]^{--}_{ii'}\Bigr]\Biggr\}, \\
\fl [\alpha_\mathrm{M}^{qq'}(\omega)]_{ii'}=& \frac{3c}{8\pi\omega}\sum_{p}\Biggl\{\int_0^{\frac{\omega}{c}}\frac{dk\,k}{k_z}   e^{ik_z (z_q-z_{q'})}  [N_p^{qq'}(k,\omega)]^{++}_{ii'}\bigl(1-|\rho_{p}(\mathbf{k},\omega)|^2 -|\tau_{p}(\mathbf{k},\omega)|^2\bigr) \\
\fl &-i\int_{\frac{\omega}{c}}^{\infty}\frac{dk\,k}{\mathrm{Im}(k_z)} e^{-\mathrm{Im}(k_z)(z_q+z_{q'})}   [N_p^{qq'}(k,\omega)]^{++}_{ii'} \bigl[\rho_{p}(\mathbf{k},\omega)-\rho_{p}(\mathbf{k},\omega)^*\bigr]\Biggr\},}\end{equation}
where, using the fact that  $\rho_{p}(\mathbf{k},\omega)$ and $\tau_{p}(\mathbf{k},\omega)$
are independent from $\theta$ (the angle formed by $\mathbf{k}$ and the $x$ axis in the plane $x y$), we have introduced the angular integrals
\begin{equation}\label{angularintegrals1}\eqalign{
&[N_p^{qq'}(k,\omega)]^{\phi\phi'}_{ii'}=\int_0^{2\pi} \frac{d\theta}{\pi} [\hat{\bbm[\epsilon]}_p^\phi(\mathbf{k},\omega)]_i
[\hat{\bbm[\epsilon]}_{p}^{\phi'}(\mathbf{k},\omega)]_{i'}^*e^{i k r_{qq'} \cos\theta}, }\end{equation}
where $r_{21}=-r_{12}$. The matrix elements different from  zero are, for $p=1$,   $[N_{1}^{qq'}]^{\phi\phi'}_{11}=\frac{2}{k r_{qq'}}J_1(k r_{qq'})$, $[N_{1}^{qq'}]^{\phi\phi'}_{22}=\frac{2}{k r_{qq'}}J_1(k r_{qq'})-2 J_2(k r_{qq'})$, while for $p=2$  are
\begin{equation}\label{angularintegrals3}\eqalign{ \fl &
[N_{2}^{qq'}]^{\phi\phi'}_{11}=   \frac{2\phi \phi'c^2 |k_z^2|}{k r_{qq'}\omega^2}\Big[J_1(k r_{qq'})- k r_{qq'} J_2(k r_{qq'})\Big], \quad [N_{2}^{qq'}]^{\phi\phi'}_{13}= -i \phi  \frac{2c^2k k_z}{\omega^2}J_1(k r_{qq'} ),\\ \fl &
[N_{2}^{qq'}]^{\phi\phi'}_{22}=  \frac{2  \phi \phi' c^2 |k_z^2| }{k r_{qq'}  \omega^2}J_1(k r_{qq'}), \quad
[N_{2}^{qq'}]^{\phi\phi'}_{31}=   -i \phi'  \frac{2c^2k k_z^{*}}{\omega^2}J_1(k r_{qq'} ), \\ \fl&
[N_{2}^{qq'}]^{\phi\phi'}_{33}= \frac{2c^2 k^2}{\omega^2}J_0(k r_{qq'} ),
}\end{equation}
where $J_n(x)$ is the n-th order Bessel function of the first kind.
For $x \to 0$, it is $J_0(x)\to 1$,  $J_1(x)\to 0$, $J_2(x)\to 0$, and $J_1(x)/x\to 1/2$, so that
$[N_{1(2)}^{qq'}(k,\omega)]^{\phi\phi'}_{ii'}$ become diagonal and reduce to the vectors defined in (55) of \cite{BellomoPRA2013} in the case of a single emitter.

To simplify the functions $[\alpha^{qq'}_\mathrm{W}(\omega)]_{ii'}$ and $[\alpha^{qq'}_\mathrm{M}(\omega)]_{ii'}$ in \eref{alphaslab} we exploit the fact that the quantities $[N_1^{qq'}(k,\omega)]^{\phi \phi'}_{ii'} $ do not depend on $\phi$ and $ \phi'$ and are real, and that in the propagative sector   $[N_2^{qq'}(k,\omega)]^{++}_{ii'}=[N_2^{qq'}(k,\omega)]^{--\,*}_{ii'}$ and $[N_2^{qq'}(k,\omega)]^{+-}_{ii'}=[N_2^{qq'}(k,\omega)]^{-+\,*}_{ii'}$. Using the  angular integrals  \eref{angularintegrals1},  equation \eref{alphaslab} can thus be rewritten as
\begin{equation}\label{alphaE and alphaM}\eqalign{
[\alpha^{qq'}_\mathrm{W}(\omega)]_{ii'}&=\frac{[A^{qq'}(\omega)]_{ii'}^*+[B^{qq'}(\omega)]_{ii'}+2[C^{qq'}(\omega)]_{ii'}}{2},\\
[\alpha^{qq'}_\mathrm{M}(\omega)]_{ii'}&=\frac{[A^{qq'}(\omega)]_{ii'}-[B^{qq'}(\omega)]_{ii'}+2[D^{qq'}(\omega)]_{ii'}}{2},
}\end{equation}
where we have introduced the integral matrices
\begin{equation}\label{integrals}\eqalign{
\fl [A^{qq'}(\omega)]_{ii'}=& \frac{3c}{4\omega} \sum_p\int_0^{\frac{\omega}{c}}\frac{k\,dk}{k_z} e^{ik_z (z_q-z_{q'})} [N_p^{qq'}(k,\omega)]^{++}_{ii'} ,\\
\fl [B^{qq'}(\omega)]_{ii'}= &\frac{3c}{4\omega}\sum_p\int_0^{\frac{\omega}{c}}\frac{k\,dk}{k_z}e^{ik_z (z_q-z_{q'})} [N_p^{qq'}(k,\omega)]^{++}_{ii'}\bigl(|\rho_{p}(k,\omega)|^2+|\tau_{p}(k,\omega)|^2\bigr),
\\
\fl [C^{qq'}(\omega)]_{ii'}=&\frac{3c}{4\omega}\sum_p\int_0^{\frac{\omega}{c}}\frac{k\,dk}{k_z} \mathrm{Re}\bigl[ e^{ik_z(z_q+z_{q'})}   [N_p^{qq'}(k,\omega)]^{+-}_{ii'} \rho_{p}(k,\omega)\bigr],\\
\fl [D^{qq'}(\omega)]_{ii'}=& \frac{3c}{4\omega}\sum_p\int_{\frac{\omega}{c}}^{+\infty}\!\!\!\!\frac{k\,dk}{\mathrm{Im}(k_z)}e^{-\mathrm{Im}(k_z)(z_q+z_{q'})} [N_p^{qq'}(k,\omega)]^{++}_{ii'}\mathrm{Im}[\rho_{p}(k,\omega)].}\end{equation}
For $q=q'$, $\alpha^{qq}_\mathrm{W}(\omega)$ and $\alpha^{qq}_\mathrm{M}(\omega)$ coincide with the functions defined in (56) of \cite{BellomoPRA2013} in the case of a single emitter. For $q \neq q'$, in the limit  $\mathbf{R}_{q'} \to \mathbf{R}_q$,
$[\alpha^{qq'}_\mathrm{W}(\omega)]_{ii'}$ and $[\alpha^{qq'}_\mathrm{M}(\omega)]_{ii'}$ tend to  their values in the case of a single emitter placed in $\mathbf{R}_q$. In the limit the distance between the two emitters goes to infinity, both $\alpha^{qq'}_\mathrm{W}(\omega)$ and $\alpha^{qq'}_\mathrm{M}(\omega)$ go to zero. In the limit of  $|r_{qq'}| \to \infty$, this is due to the fact that the functions $[N_p^{qq'}(k,\omega)]^{\phi\phi'}$  go to zero (for $|x| \to \infty$, it is $J_0(x)\to 0$,  $J_1(x)\to 0$ and $J_2(x)\to 0$). In the limit $|z_q-z_{q'}|\to \infty$, this is due to the presence of an oscillating functions whose frequency goes to infinity in the integrals $A(\omega)$, $B(\omega)$ and $C(\omega)$ (this can be seen explicitly by integrating by parts) and to the presence of the exponential function going to zero in the integral $D(\omega)$.

Concerning the function $\Lambda^{qq'}(\omega)$, in order to develop its expression in \eref{lambda3} one has to compute the real part of the Green's function in terms of the scattering operators $\mathcal{R}$ and $\mathcal{T}$. This is done in  \ref{par:Green function} where a free term $\Lambda_0^{qq'}(\omega)$ [see  \eref{freeLambda}] remaining in absence of matter has been isolated from a reflected part $\Lambda_R^{qq'}(\omega)$ [see  \eref{ReGR}], $\Lambda^{qq'}(\omega)=\Lambda_0^{qq'}(\omega)+\Lambda_R^{qq'}(\omega)$.
Using the expressions for  $\mathcal{R}$ and $\mathcal{T}$ in the case of a slab and the  angular integrals in \eref{angularintegrals1}, one can derive, starting from   \eref{ReGR}:
\begin{equation}\label{alphaslab2}\eqalign{ \fl &\mathrm{Re} \, G^{(R)}_{ii'}(\mathbf{R}_q,\mathbf{R}_{q'},\omega)=\frac{i \omega^2}{4\epsilon_0c^2}\frac{1}{4 \pi}\sum_{p}\Biggl\{\int_0^{\frac{\omega}{c}}\frac{k\, dk}{k_z}
\Bigl[e^{ik_z(z_q+z_{q'})} [N_p^{qq'}(k,\omega)]^{+-}_{ii'}\rho_{p}(\mathbf{k},\omega)\\
\fl &\qquad-e^{-ik_z(z_q+z_{q'})}[N_p^{qq'}(k,\omega)]^{-+}_{ii'}\rho_{p}(\mathbf{k},\omega)^*-i\int_{\frac{\omega}{c}}^{\infty}\frac{dk\,k}{\mathrm{Im}(k_z)}e^{-\mathrm{Im}(k_z)(z_q+z_{q'})}\\
\fl &\qquad\times [N_p^{qq'}(k,\omega)]^{++}_{ii'} \bigl(\rho_{p}(\mathbf{k},\omega)+\rho_{p}(\mathbf{k},\omega)^*\bigr)\Bigr]\Biggr\}.\\
}\end{equation}
Equation \eref{lambda3} can be thus cast under the form
\begin{equation}\label{lambda4}\eqalign{
\fl \Lambda^{qq'}(\omega)= &\Lambda_0^{qq'}(\omega)+\sqrt{\Gamma_0^q(\omega)\Gamma_0^{q'}(\omega) }  \sum_{i,i'}[\tilde{\textbf{d}}_{mn}^q]^*_{i} [\tilde{\textbf{d}}_{m'n'}^{q'}]_{i'}
\bigr([C_2^{qq'}(\omega)]_{ii'}-[D_2^{qq'}(\omega)]_{ii'}\bigr),
 }
 \end{equation}
where we have introduced the integral matrices
\begin{equation}\label{integrals2}\eqalign{
\fl [C_2^{qq'}(\omega)]_{ii'}=&\frac{3c}{8\omega}\sum_p\int_0^{\frac{\omega}{c}}\frac{k\,dk}{k_z} \mathrm{Im}\bigl[ e^{ik_z(z_q+z_{q'})}  [N_p^{qq'}(k,\omega)]^{+-}_{ii'}  \rho_{p}(k,\omega)\bigr],\\
\fl [D_2^{qq'}(\omega)]_{ii'}=& \frac{3c}{8\omega}\sum_p\int_{\frac{\omega}{c}}^{+\infty}\!\!\!\!\frac{k\,dk}{\mathrm{Im}(k_z)}e^{-\mathrm{Im}(k_z)(z_q+z_{q'})} [N_p^{qq'}(k,\omega)]^{++}_{ii'}\mathrm{Re}\bigl[\rho_{p}(k,\omega)\bigr].\\}\end{equation}

We observe that  the limit case when the body is absent is discussed in  \ref{app:absence of matter}, where known expressions for $\Gamma^{qq'}(\omega)$ and $\Lambda^{qq'}(\omega)$ are retrieved.

\section{Two-qubit system \label{par:twoqubits}}

From now on we specialize our investigation to the case of two emitters (qubits) characterized by two internal levels
  $\ket{1}\equiv \ket{g}$ and
$\ket{2 }\equiv \ket{e}$ with the same transition frequency $\omega=\omega_e^1-\omega_g^1=\omega_e^2-\omega_g^2$.
In this case,  master equation  \eref{master equation 5} reduces to
\begin{equation}\label{master equation 6} \eqalign{ \fl \frac{d}{d t}\rho =& - \frac{i}{\hbar} [H_S+  \delta_S,\rho ]
-i\sum_{q\neq q'}\Lambda^{qq'}(\omega)[\sigma_{ge}^{q\,\dag}\sigma_{ge}^{q'},\rho ]
+\sum_{q, q'}\Gamma^{qq'}(\omega)\Big(\sigma^{q'}_{ge}\rho \sigma_{ge}^{q\,\dag}\\ \fl & -\frac{1}{2}\{\sigma^{q\,\dag}_{ge}\sigma^{q'}_{ge},\rho \}\Big) +\sum_{q, q'}\Gamma^{qq'}(-\omega)\Big(\sigma^{q'\dag}_{ge}\rho \sigma^{q}_{ge}-\frac{1}{2}\{\sigma^{q}_{ge} \sigma^{q'\,\dag}_{ge},\rho \}\Big),}\end{equation}
where we used $[\sigma_{gg}^q,\rho ]=-[\sigma_{ee}^q,\rho ]$, so that
\begin{equation}\label{shift operator}
\delta_S=\sum_{q}\hbar  \bigl[S^{qq}(\omega)-S^{qq}(-\omega)\bigr] \sigma_{ee}^q,
\end{equation}
and where the functions $S^{qq}(\pm\omega)$, $\Lambda^{qq'}(\omega)$ and $\Gamma^{qq'}(\pm\omega)$ are defined in  \eref{me parameters N} for the specific case $\{m,n\}=\{1,2\}$ (in the following we use the notation $\mathbf{d}_{12}^q=\mathbf{d}^{q}$).

We observe that master equation \eref{master equation 6} can also describe the case in which the emitters’ frequencies are close enough (but not identical) so that rotating wave approximation used in the derivation of\eref{master equation} still holds. This typically occurs when the frequency difference is much smaller
than the average frequency \cite{FicekBook2005}.
The operator $\delta_S$ \eref{shift operator} represents a shift of energy levels, being the renormalized transition frequencies equal to
 $\tilde{\omega}_e^1-\omega_g^1=\omega+S^{11}(\omega) -S^{11}(-\omega)$ and $\tilde{\omega}_e^2-\omega_g^2= \omega+S^{22}(\omega) -S^{22}(-\omega)$.
  {When these shifts are equal among them, that is $S^{11}(\omega) -S^{11}(-\omega)= S^{22}(\omega) -S^{22}(-\omega)$, as in the case of two identical qubits placed at the same distance from a slab, they do not play any relevant role in the dynamics. This is not the case in general, if the two shifts are not equal. However, in the cases treated in the following, we obtained numerical evidence that when they are not equal their influence is small and it will then be neglected.}

To discuss the properties of \eref{master equation 6}, we will use two different bases,
the decoupled basis  $\{\ket{1}\equiv \ket{g g},\ket{2}\equiv \ket{e g},\ket{3}\equiv \ket{g e},\ket{4}\equiv \ket{e e}\}$ and the coupled basis $\{\ket{\mathrm{G}}\equiv \ket{1},
\ket{\mathrm{A}}\equiv (\ket{2}-\ket{3})/\sqrt{2},\ket{\mathrm{S}}\equiv(\ket{2}+\ket{3})/\sqrt{2},\ket{\mathrm{E}}\equiv \ket{4}\}$, where we have introduced the collective antisymmetric $\ket{\mathrm{A}}$ and symmetric states $\ket{\mathrm{S}}$.
The coupled basis is the one diagonalizing the effective Hamiltonian,  {$H_S+\sum_{q\neq q'}\hbar \Lambda^{qq'}(\omega)\sigma_{ge}^{q\,\dag}\sigma_{ge}^{q'}$}, appearing in the first line of  \eref{master equation 6}. In particular, the sign of $\Lambda^{12}(\omega)$ inverts the role of $\ket{\mathrm{A}}$ and $\ket{\mathrm{S}}$ in the eigenstates of the above effective Hamiltonian. The eigenvalues associated to  $\ket{\mathrm{G}}$, $\ket{\mathrm{A}}$, $\ket{\mathrm{S}}$, $\ket{\mathrm{E}}$, are   {$\{0,$ $ \hbar(\omega- |\Lambda^{12}(\omega)|), \hbar(\omega+ |\Lambda^{12}(\omega)|), 2 \hbar \omega \}$, having set the energy of the ground state equal to zero}.

\subsection{X states}

In the  decoupled basis we can distinguish elements along the two main diagonals of the two-qubit density matrix from the remaining ones because they are not connected through master equation \eref{master equation 6}. We thus focus our attention on the class of X states, having non-zero elements only along the main diagonal and
anti-diagonal of the density matrix (we use the notation $\rho_{ij}=\bra{i}\rho \ket{j}$),
\begin{equation}\label{Xstates}
   \rho_X = \left(
\begin{array}{cccc}
  \rho_{11} & 0 & 0 & \rho_{14}  \\
  0 & \rho_{22} & \rho_{23} & 0 \\
  0 & \rho_{23}^* & \rho_{33} & 0 \\
  \rho_{14} ^* & 0 & 0 & \rho_{44} \\
\end{array}
\right).
\end{equation}
Bell, Werner  and Bell diagonal states belong to this class of states
\cite{BellomoASL}.  X-structure density matrices are found  in a wide variety of physical situations and are also experimentally achievable \cite{Pratt2004PRL}. For example, X states are encountered as eigenstates in all the systems with odd-even symmetry like in the Ising and the XY models \cite{Fazio2002Nature}. Moreover, in many physical evolutions of open quantum systems an initial X structure is maintained in time \cite{Yu2007},
as it is in our case. Terms outside the two main diagonals initially populated, would be eventually washed off asymptotically. In the following, the two-qubit state will have always an X structure.

\subsection{Concurrence}

We shall quantify the entanglement in the two-qubit dynamics by evaluating the
concurrence, $C(t)$
($C=0$ for separable states, $C=1$ for  maximally entangled states)~\cite{wootters1998PRL}.
For  X states it takes the form~\cite{Yu2007}
\begin{equation}\label{concxstate}\eqalign{
C(t)&=2\,\mathrm{max}\{0,K_1(t),K_2(t)\},   \\
K_1(t)&=|\rho_{23}(t)|-\sqrt{\rho_{11}(t)\rho_{44}(t)}, \quad
K_2(t)=|\rho_{14}(t)|-\sqrt{\rho_{22}(t)\rho_{33}(t)}. }
\end{equation}

The master equation \eref{master equation 6} always induces an exponential decay for $\rho_{14}(t)$, so that in the steady state only $K_1(t)$ could be responsible for having $C(\infty)>0$.

To discuss new phenomena emerging out of thermal equilibrium, it will be instructive to rewrite  $K_1(t)$ in terms of the populations in the coupled basis (we use the notation $\rho_{\mathrm{IJ}}=\bra{\mathrm{I}}\rho \ket{\mathrm{J}}$ and $\rho_{\mathrm{I}}=\bra{\mathrm{I}}\rho \ket{\mathrm{I}}$):
\begin{equation}\eqalign{\label{conccoupled}
K_1(t)=&\frac{1}{2}\sqrt{[\rho_{\mathrm{S}}(t)-\rho_{\mathrm{A}}(t)]^2+ {|}\rho_{\mathrm{SA}}(t)-\rho_{\mathrm{AS}}(t) {|}^2}  -\sqrt{\rho_{\mathrm{G}}(t)\rho_{\mathrm{E}}(t)}.
}\end{equation}
We will see that out of thermal equilibrium, it is always $\rho_{\mathrm{AS}}(\infty)=0$,  but $\rho_{\mathrm{S}}(\infty)$ and $\rho_{\mathrm{A}}(\infty)$ can differ, so that $K_1(\infty)$ could be positive.

\subsection{Thermal equilibrium}

When $T_\mathrm{W}=T_\mathrm{M}\equiv T$, master equation \eref{master equation 6} describes the thermalization towards the thermal equilibrium state, which is diagonal with the four steady populations given by
\begin{equation}\label{thermal state}
  \left(
\begin{array}{c}
 \rho_{11}(\infty)\\
    \rho_{22}(\infty)\\
  \rho_{33}(\infty)\\
    \rho_{44}(\infty)\\
\end{array}
\right)_{\!\!\!\mathrm{eq}} =\frac{1}{Z_{\mathrm{eq}}}\left(
\begin{array}{c}
 [1+ n(\omega, T)]^2 \\
   n(\omega, T)[1+ n(\omega, T)] \\
 n(\omega, T)[1+ n(\omega, T)]\\
   n(\omega, T)^2\\
\end{array}
\right),
\end{equation}
where $Z_{\mathrm{eq}}=[1+2 n(\omega, T)]^2$. By moving to the coupled basis, the thermal state remains diagonal with $ \rho_{\mathrm{S}}(\infty)= \rho_{\mathrm{A}}(\infty)= \rho_{22}(\infty)= \rho_{33}(\infty)$.

As a mathematical remark, we note that the thermal state is always reached asymptotically except if the identities $\Gamma^{11}(\pm\omega)=\Gamma^{22}(\pm\omega)=\Gamma^{12}(\pm\omega)=\Gamma^{21}(\pm\omega)\equiv \Gamma(\pm\omega)$ are strictly verified. In this peculiar case, both in and out of thermal equilibrium, the steady state depends upon the initial state and may be entangled. In particular, it is diagonal in the coupled basis with populations equal to
 \begin{equation}\label{singular case}
   \left(
\begin{array}{c}
 \rho_{\mathrm{G}}(\infty)\\
    \rho_{\mathrm{A}}(\infty)\\
  \rho_{\mathrm{S}}(\infty)\\
    \rho_{\mathrm{E}}(\infty)\\
\end{array}
\right)_{}  =\frac{1}{Z}\left(
\begin{array}{c}
\Gamma(\omega)^2 [1- \rho_{\mathrm{A}} (0)] \\
\rho_{\mathrm{A}} (0)\\
  \Gamma(-\omega)\Gamma(\omega) [1- \rho_{\mathrm{A}} (0)]\\
  \Gamma(-\omega)^2 [1- \rho_{\mathrm{A}} (0)]
\end{array}
\right),
\end{equation}
where $Z=\Gamma(-\omega)^2+\Gamma(\omega)\Gamma(-\omega)+\Gamma(\omega)^2$.
Apart from this case, at thermal equilibrium the steady state is always a thermal state, thus not entangled.
We can see it by looking at the concurrence  \eref{concxstate} which is zero
 being $\rho_{23}(\infty)=0$.
This can also be seen in  the coupled basis,  where $\rho_{\mathrm{AS}}(\infty)=0$ and $\rho_{\mathrm{S}}(\infty)=\rho_{\mathrm{A}}(\infty)$, so that $K_1(\infty)$  \eref{conccoupled} is  negative.

\subsection{Out of thermal equilibrium: an instructive case \label{subpar:analytic investigation}}

\begin{figure}[t!]
\begin{center} \includegraphics[width=0.55\textwidth]{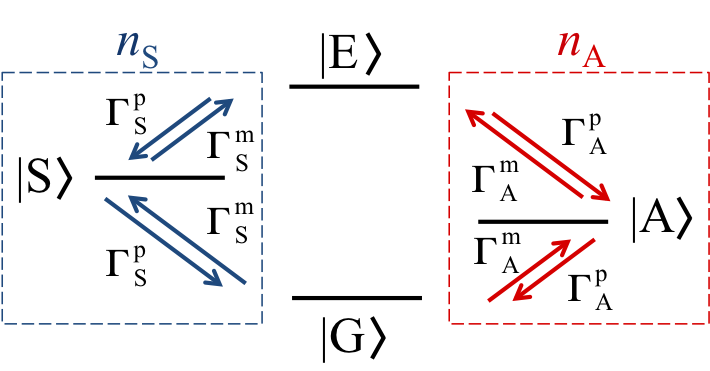} \end{center}
\caption{\label{fig:figuraschema}\footnotesize Representation of the rate equations  \eref{MEsymm}. The transition rates in the two channels are  $\Gamma^{\mathrm{p}}_{\mathrm{S}(\mathrm{A})}=\Gamma_{\mathrm{S}(\mathrm{A})}(1+ n_{\mathrm{S}(\mathrm{A})})$ and $\Gamma^{\mathrm{m}}_{\mathrm{S}(\mathrm{A})}=\Gamma_{\mathrm{S}(\mathrm{A})}\, n_{\mathrm{S}(\mathrm{A})}.$}
\end{figure}

When $T_\mathrm{W}\neq T_\mathrm{M}$, qualitative differences emerge in the dynamics and in the steady states. To highlight these new features,
we first consider a simple case where a clear physical interpretation in terms of $\ket{S}$ and $\ket{A}$ is available. This is the case when $\Gamma^{11}(\pm\omega)=\Gamma^{22}(\pm\omega)\equiv \Gamma(\pm\omega)$ and $\Gamma^{12(21)}(\pm\omega) $ are real. These conditions are verified, for example, in the case of identical qubits, with $\mathbf{d}^1=\mathbf{d}^2\equiv\mathbf{d} $, placed in equivalent positions with respect to the body (in the case of a slab, $z_1=z_2$) and with $\mathbf{d}$ real and having components different from zero either only along the $z$ axis or only along the plane $x y$. In this case,  master equation \eref{master equation 6} gives in the coupled basis  a set of rate equations for the populations, which are decoupled from the other density matrix elements:
\begin{equation}\label{MEsymm}\label{MEsymm}\label{MEsymm}\eqalign{
\dot{\rho}_{\mathrm{G}}=&-(\Gamma_\mathrm{A} \,n_\mathrm{A}+\Gamma_\mathrm{S}\, n_\mathrm{S})\rho_{\mathrm{G}}+\Gamma_\mathrm{A}(1+n_\mathrm{A}) \rho_{\mathrm{A}}+\Gamma_\mathrm{S}(1+n_\mathrm{S}) \rho_{\mathrm{S}},\\
\dot{\rho}_{\mathrm{A}}=&-\Gamma_\mathrm{A} (1+2 n_\mathrm{A})  \rho_{\mathrm{A}}+\Gamma_\mathrm{A}\, n_\mathrm{A} \rho_{\mathrm{G}}+\Gamma_\mathrm{A}(1+n_\mathrm{A}) \rho_{\mathrm{E}},\\
\dot{\rho}_{\mathrm{S}}=&-\Gamma_\mathrm{S} (1+2 n_\mathrm{S})  \rho_{\mathrm{S}}+\Gamma_\mathrm{S}\, n_\mathrm{S} \rho_{\mathrm{G}}+\Gamma_\mathrm{S}(1+n_\mathrm{S}) \rho_{\mathrm{E}},\\
\dot{\rho}_{\mathrm{E}}=&-[\Gamma_\mathrm{A}  {(1+ n_\mathrm{A})} +\Gamma_\mathrm{S}  {(1+ n_\mathrm{S})} ] \rho_{\mathrm{E}}+\Gamma_\mathrm{A}\, n_\mathrm{A} \rho_{\mathrm{A}}  +\Gamma_\mathrm{S}\, n_\mathrm{S} \rho_{\mathrm{S}}.
}\end{equation}
Here the coefficient $\Gamma_0(\omega)$ has been absorbed by the time variable in the derivative, which is now dimensionless, and we have used the relations
\begin{equation} \label{channelrates} \eqalign{
\fl \Gamma(\omega)-\Gamma^{12}(\omega) =\Gamma_0(\omega) \Gamma_\mathrm{A} (1+n_\mathrm{A}), \quad \Gamma(\omega)+\Gamma^{12}(\omega)=\Gamma_0(\omega) \Gamma_\mathrm{S} (1+n_\mathrm{S}),\\
\fl \Gamma(-\omega)-\Gamma^{12}(-\omega) =\Gamma_0(\omega) \Gamma_\mathrm{A} \,n_\mathrm{A},\quad \Gamma(-\omega)+\Gamma^{12}(-\omega)=\Gamma_0(\omega) \Gamma_\mathrm{S} \,n_\mathrm{S},
}\end{equation}
with \begin{equation}\eqalign{\label{nA e nS}
\fl \Gamma_\mathrm{A}=\alpha_\mathrm{W}(\omega) -\alpha^{12}_\mathrm{W}(\omega)+\alpha_\mathrm{M}(\omega) -\alpha^{12}_\mathrm{M}(\omega),\\
\fl  \Gamma_\mathrm{S}=\alpha_\mathrm{W}(\omega) +\alpha^{12}_\mathrm{W}(\omega)+\alpha_\mathrm{M}(\omega) +\alpha^{12}_\mathrm{M}(\omega),\\
\fl  n_\mathrm{A} =\frac{1}{\Gamma_\mathrm{A}}\Bigl\{ \left[\alpha_\mathrm{W}(\omega) -\alpha^{12}_\mathrm{W}(\omega)\right] n(\omega,T_\mathrm{W})  +\left[\alpha_\mathrm{M}(\omega) -\alpha^{12}_\mathrm{M}(\omega)\right]n(\omega,T_\mathrm{M}) \Bigr\},\\
\fl n_\mathrm{S}  = \frac{1}{\Gamma_\mathrm{S}}\Bigl\{ \left[\alpha_\mathrm{W}(\omega) +\alpha^{12}_\mathrm{W}(\omega)\right] n(\omega,T_\mathrm{W})
 +\left[\alpha_\mathrm{M}(\omega) +\alpha^{12}_\mathrm{M}(\omega)\right]n(\omega,T_\mathrm{M}) \Bigr\},
}\end{equation}
where $\alpha_\mathrm{W(M)}(\omega) \equiv \alpha^{11}_\mathrm{W(M)}(\omega)= \alpha^{22}_\mathrm{W(M)}(\omega)$. We remark that $\Lambda^{qq'}(\omega)$ does not enter in the rate equations  \eref{MEsymm},
which are schematically represented in figure \ref{fig:figuraschema}. We observe that to each decay channel from $\ket{\mathrm{E}}$ to $\ket{\mathrm{G}}$ we can associate distinct effective temperatures $T_\mathrm{S}$ and $T_\mathrm{A}$ confined between $T_\mathrm{W}$ and $T_\mathrm{M}$  in correspondence to the effective number of photons $n_\mathrm{S}$ and $n_\mathrm{A}$, which have the property of  being confined between $n(\omega,T_\mathrm{W})$ and $n(\omega,T_\mathrm{M})$ \cite{BellomoPRA2013}.
Concerning the coherences in the second diagonal:
\begin{equation}\eqalign{\label{MEsymm coher}
\dot{\rho}_{\mathrm{A}\mathrm{S}}=&-\frac{1}{2}\left[\Gamma_\mathrm{A} (1+2 n_\mathrm{A})  +\Gamma_\mathrm{S}(1+2 n_\mathrm{S}) -4 i\Lambda^{12}(\omega)\right] \rho_{\mathrm{A}\mathrm{S}},\\
\dot{\rho}_{\mathrm{G}\mathrm{E}}=&-\frac{1}{2}\left[\Gamma_\mathrm{A} (1+2 n_\mathrm{A})  +\Gamma_\mathrm{S}(1+2 n_\mathrm{S})-4 i  {\omega} \right] \rho_{\mathrm{G}\mathrm{E}},\\
}\end{equation}
which give  for each coherence an exponential decay modulating oscillations due {, respectively,} to $\Lambda^{12}(\omega)$  {and $\omega$}.
The stationary solution of  \eref{MEsymm}  is
\begin{equation}\eqalign{\label{ote case} \fl
&   \left(
\begin{array}{c}
 \rho_{\mathrm{G}}(\infty)\\
 \\
    \rho_{\mathrm{A}}(\infty)\\
    \\
  \rho_{\mathrm{S}}(\infty)\\
  \\
    \rho_{\mathrm{E}}(\infty)\\
\end{array}
\right)_{\!\!\!\mathrm{neq}}  =\frac{1}{Z_{\mathrm{neq}}}
 \left(
\begin{array}{c}
(1+n_\mathrm{A})^2 (1+2n_\mathrm{S})\Gamma_\mathrm{A}+(1+2 n_\mathrm{A})(1+n_\mathrm{S})^2\Gamma_\mathrm{S}\\ \\
n_\mathrm{A}(1+n_\mathrm{A})(1+2n_\mathrm{S}) \Gamma_\mathrm{A}+[ n_\mathrm{A}(1+2n_\mathrm{S}) \\+n_\mathrm{S}^2(1+2n_\mathrm{A}) ] \Gamma_\mathrm{S} \\\\
n_\mathrm{S}(1+n_\mathrm{S})(1+2n_\mathrm{A}) \Gamma_\mathrm{S}+[ n_\mathrm{S}(1+2n_\mathrm{A}) \\+n_\mathrm{A}^2(1+2n_\mathrm{S}) ] \Gamma_\mathrm{A} \\ \\
n_\mathrm{A}^2 (1+2n_\mathrm{S})\Gamma_\mathrm{A}+(1+2 n_\mathrm{A})n_\mathrm{S}^2\Gamma_\mathrm{S}
\end{array}
\right),
}\end{equation}
where $Z_{\mathrm{neq}}$ is the sum of the elements of the vector in the second line of the above equation. Out of  equilibrium $  \rho_{23}(\infty)_{\mathrm{neq}}$ is different from zero and is given by:
\begin{equation}\eqalign{\label{rho23ote}
  \rho_{23}(\infty)_{\mathrm{neq}} =\frac{(n_\mathrm{S}-n_\mathrm{A})(\Gamma_\mathrm{S}+\Gamma_\mathrm{A})}{2 Z_{\mathrm{neq}}},
}\end{equation}
where we see easily how it tends to zero at thermal equilibrium when $n_\mathrm{S}=n_\mathrm{A}$.
Using \eref{ote case} in  \eref{concxstate} and \eref{conccoupled}, we obtain for the steady concurrence:
\begin{equation}\eqalign{\label{Cote}
 C(\infty)_{\mathrm{neq}}=& \frac{2}{Z_{\mathrm{neq}}} \Bigl[ |n_\mathrm{S}-n_\mathrm{A}|(\Gamma_\mathrm{S}+\Gamma_\mathrm{A})  /2  \\
 &\,-\sqrt{(1+n_\mathrm{A})^2 (1+2n_\mathrm{S})\Gamma_\mathrm{A}+(1+2 n_\mathrm{A})(1+n_\mathrm{S})^2\Gamma_\mathrm{S}}\\
 &\, \times\sqrt{n_\mathrm{A}^2 (1+2n_\mathrm{S})\Gamma_\mathrm{A}+(1+2 n_\mathrm{A})n_\mathrm{S}^2\Gamma_\mathrm{S}} \; \Bigr] .
}\end{equation}
\begin{figure}[t!]
\begin{center} \includegraphics[width=0.65\textwidth]{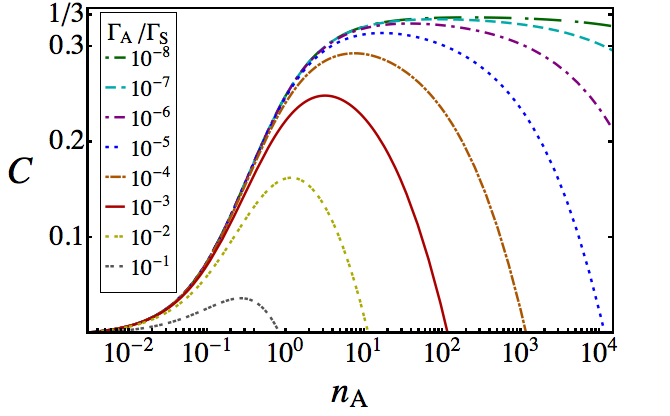} \end{center}
\caption{\label{fig:figuraconcorrenza}\footnotesize  $C= C(\infty)_{\mathrm{neq}}$ vs $n_{\mathrm{A}}$ for $n_{\mathrm{S}}=10^{-3}$ for different values of $\Gamma_{\mathrm{A}}/\Gamma_{\mathrm{S}}$ indicated in the legend.}
\end{figure}
Simplifying  $\Gamma_\mathrm{S}$,  $C(\infty)_{\mathrm{neq}}$ becomes function of the three dimensionless quantities $\Gamma_\mathrm{A}/\Gamma_\mathrm{S}$, $n_\mathrm{S}$ and $n_\mathrm{A}$. This dependence is discussed in figure \ref{fig:figuraconcorrenza}, where $C=C(\infty)_{\mathrm{neq}}$ is depicted as a function of  $n_\mathrm{A}$ for  $n_\mathrm{S}=0.001$ and for several values of $\Gamma_\mathrm{A}/\Gamma_\mathrm{S}$, as indicated in the legend. We observe that by decreasing the  value of $\Gamma_\mathrm{A}/\Gamma_\mathrm{S}$, higher values of $C$ are reachable at higher values of  $n_\mathrm{A}$. The maximum value of $C$ is 1/3, which can be obtained in the limits $\Gamma_\mathrm{A}/\Gamma_\mathrm{S} \to 0, n_\mathrm{S} \to 0$ and $n_\mathrm{A} \to \infty $. In particular, the corresponding maximally entangled state, which is a statistical mixture of
the ground and of the antisymmetric state
with weights respectively equal to 2/3 and 1/3, has
also been found in \cite{Camalet2013}. For smaller values of $n_\mathrm{S}$ the behavior remains almost identical, while by increasing its value, the values of $C$ decrease progressively. We remark that an identical behavior is found in the opposite case, i.e.  when $\Gamma_\mathrm{S}/\Gamma_\mathrm{A} \to 0$, case in which the role of states $\ket{\mathrm{S}}$ and $\ket{\mathrm{A}}$ is inverted. This can be achieved by looking for values of the various parameters such that $\alpha^{12}_\mathrm{M(W)}(\omega)$ is negative and very close to $\alpha_\mathrm{M(W)}(\omega)$ in order make the ratio $\Gamma_\mathrm{S}/\Gamma_\mathrm{A} $ very small.

Figure \ref{fig:figuraconcorrenza} describes the generation of steady entangled states emerging only in the absence of thermal equilibrium. Two main conditions must be fulfilled, the first being to have small values  for $\Gamma_\mathrm{A}/\Gamma_\mathrm{S}$ and the second to realize (quite) different effective temperatures for the two decay channels, which can be achieved only in absence of thermal equilibrium.

\section{Numerical investigation \label{par:numerical investigation}}

Here we report the numerical investigation concerning the case treated in Sec. \ref{par: atom in front of a slab} when the body close to the emitters is  a slab of finite thickness $\delta$. According to \eref{master equation}, a relevant parameter involved in our investigation concerning the role of the body is the value of the dielectric permittivity at the common transition frequency of the two qubits.
As material we choose the silicon carbide (SiC) whose dielectric permittivity $\varepsilon (\omega)$  is described using a Drude-Lorentz model \cite{Palik98}
\begin{equation}\varepsilon (\omega) =\varepsilon_{\infty}\frac{\omega ^2-\omega_l^2+i\Gamma\omega}{\omega^2-\omega_r^2+i\Gamma\omega},\end{equation}
characterized by a resonance at $\omega_r=1.495\times10^{14}\,\mathrm{rad}\,\mathrm{s}^{-1}$ and where $\varepsilon_{\infty}=6.7 $, $\omega_l=1.827\times10^{14}\,\mathrm{rad}\,\mathrm{s}^{-1}$ and $\Gamma=0.009\times10^{14}\,\mathrm{rad}\,\mathrm{s}^{-1}$. This model implies a surface phonon-polariton resonance at $\omega_p=1.787\times10^{14}\,\mathrm{rad}\,\mathrm{s}^{-1}$. A relevant length scale in this case is $c/\omega_r\simeq 2\,\mu$m while a reference temperature is $\hbar\omega_r/k_B\simeq 1140\,$K. We will assume that $\varepsilon (\omega)$ does not vary much in the interval of temperatures considered. In the following study, we explore a region of parameters much wider than that  allowing the analytical description of Sec. \ref{subpar:analytic investigation}.

\subsection{Steady configurations}

We first focus on the properties of  steady states. In particular, we are interested in the amount of entanglement present asymptotically, which is quantified by the concurrence  \eref{concxstate}. This analysis is supported by an analytical solution of the steady state of  \eref{master equation 6} which is not reported here, since particularly cumbersome.

\begin{figure}[t!]
\begin{center}\includegraphics[width=0.96\textwidth]{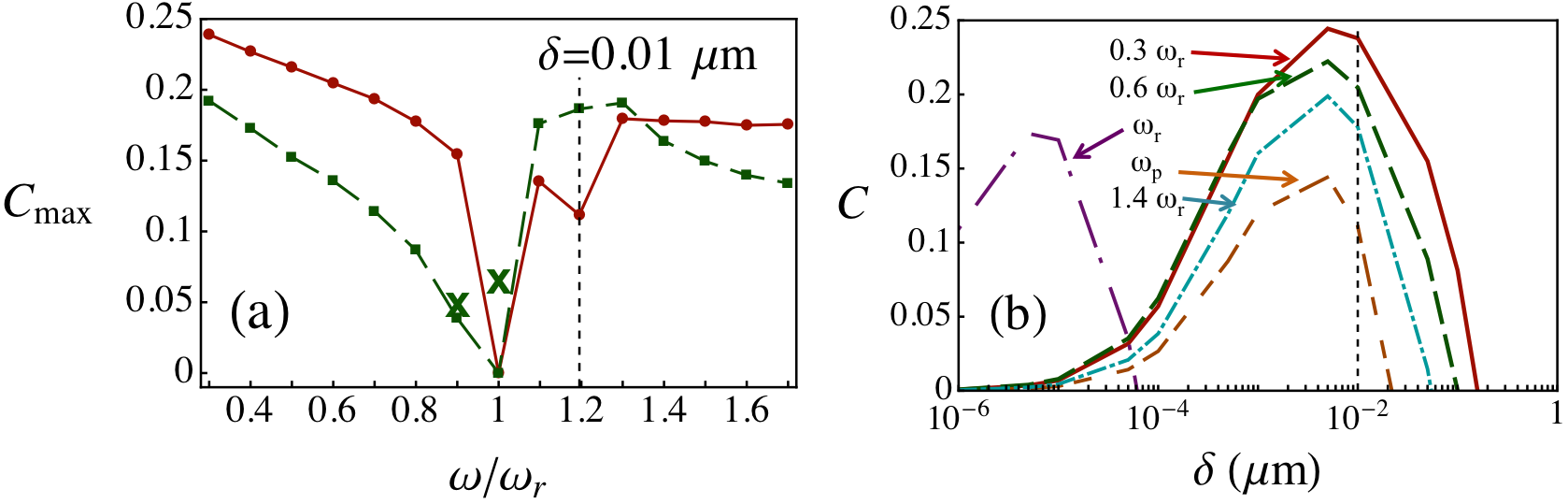}\end{center}
\caption{\label{fig:FigCmax}\footnotesize  Part (a): $C_{\mathrm{max}}$ vs $\omega/\omega_r$ for $\delta=0.01 \mu$m. The red curve concerns the case of identical electric dipoles oriented along the $z$ axis and the green along the $x$ axis. The solid (red) and dashed (green) lines just connect the sampled frequencies. Crosses indicate the occurrence of larger values of $C$ at the same frequency but for different values of $\delta$. The black dotted vertical line concerns the frequency $\omega_p \simeq 1.2 \omega_r$.  Part (b): $C$ vs $\delta$ for several values of $\omega$ indicated in the figure in the case of identical electric
dipoles oriented along the z-axis.}
\end{figure}

\begin{figure}[t!]
\begin{center} \includegraphics[width=1\textwidth]{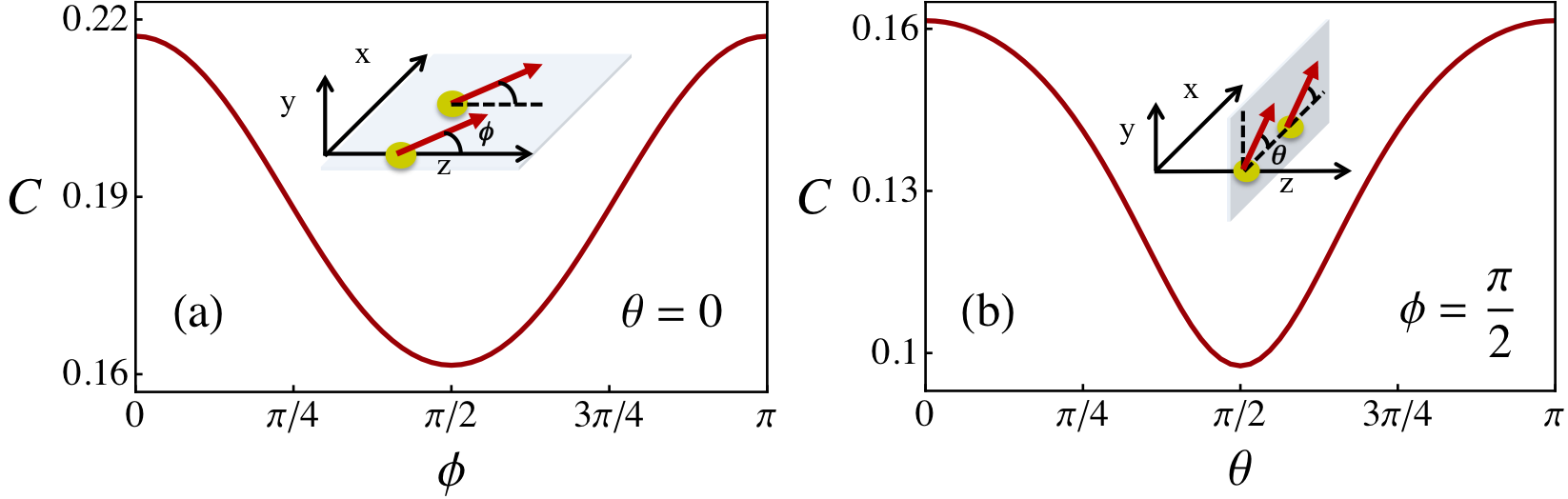} \end{center}
\caption{\label{fig:FigAng2}\footnotesize   Part (a): $C$ vs $\phi$, angle formed between the dipole directions and the $z$ axis (see inset), in the case $\theta=0$ ($\theta$ is the angle between the projection of dipole directions in the $xy$ plane and the $x$ axis). Part (b): $C$ vs $\theta$ (see inset) in the case $\phi=\pi/2$ (dipole directions in the $xy$ plane). Values of parameters:  $T_{\mathrm{W}}= 30 $ K, $T_{\mathrm{M}}= 1200 $ K, $\delta=0.01 \mu$m, $z=z_1=z_2 {\simeq 1.04} \,\mu$m and  $r_{12}$=0.25 $\mu$m. }
\end{figure}

In figure \ref{fig:FigCmax} (a), we plot the maximum of steady concurrence obtained for an interval of transition frequencies ranging from $0.3\, \omega_r$ to $1.7 \, \omega_r$, in the case of $\delta=0.01 \mu$m. In our numerical sample, $z_1$ and $z_2$ may vary between 0.05 and 50 $\mu$m and $r_{12}$ between 0 and 15 $\mu$m. The two temperatures range in an interval such that the associated number of photons is between 0 and 3 (we checked that larger values are not needed).  The red curve is relative to the case of dipoles oriented along the z axis, while the green curve to the case of dipoles oriented along the $x$ axis [in figure \ref{fig:FigAng2} we will show that this is the best choice if we limit ourselves to directions lying in the $xy$ plane]. Higher values of concurrence are obtained immediately before/after the resonance frequency $\omega_r$. The red configuration gives always better results except around the surface phonon-polariton frequency $\omega_p \simeq 1.2 \omega_r$ where choosing the dipole directions along the $x$ axis is the best choice.
The values of the parameters corresponding to each maximum vary with frequency. The best configuration is always characterized by values of $n(\omega,T_\mathrm{W})$ close to zero and  $n(\omega,T_\mathrm{M})$ between 1 and 3. Smaller values of $n(\omega,T_\mathrm{M})$ are needed in the green curve.  The zone where to place the qubits is around 1 $\mu$m from the slab at $0.3 \, \omega_r$, gradually decreasing (specially after $\omega_r$) down to 0.25  $\mu$m at $1.7  \,\omega_r$. For the red curve the best choice is always $z_1$ close to $z_2$ (in our numerical sample, we limit the minimal distance at the order of  0.1 $\mu$m) and $r_{12}=0$, while for the green curve it is $z_1=z_2$ and $r_{12}$ small (of the order of 0.01 $\mu$m). This means that the best configuration is when the interatomic axis is aligned with dipoles direction. For $\omega$ around $\omega_p$ we point out the occurrence of larger values of $C$ for different values of $\delta$, points indicated with a cross above the green curve. In the absence of large values of $C$ in correspondence to the canonical choice of the parameters described above, small values of $C$ become evident for a different set of parameters. This corresponds to larger values of $\delta$ (of the order of $1 \mu$m or more),  $z_1\simeq 2 \mu$m, $z_2\simeq 4 \mu$m and $r_{12}\simeq 0.5 \mu$m. In part (b), we plot the dependence of $C$ on $\delta$ for several values of $\omega$ as indicated in the figure. The maximum of $C$ is always obtained close to $\delta=0.01 \mu$m, which is the value chosen in part (a), except around $\omega_r$ where much smaller values of $\delta$ are required. This explains why in part (a) concurrence decreases around $\omega_r$ for $\delta=0.01 \mu$m.

 \begin{figure}[t!]
\begin{center} \includegraphics[width=0.7\textwidth]{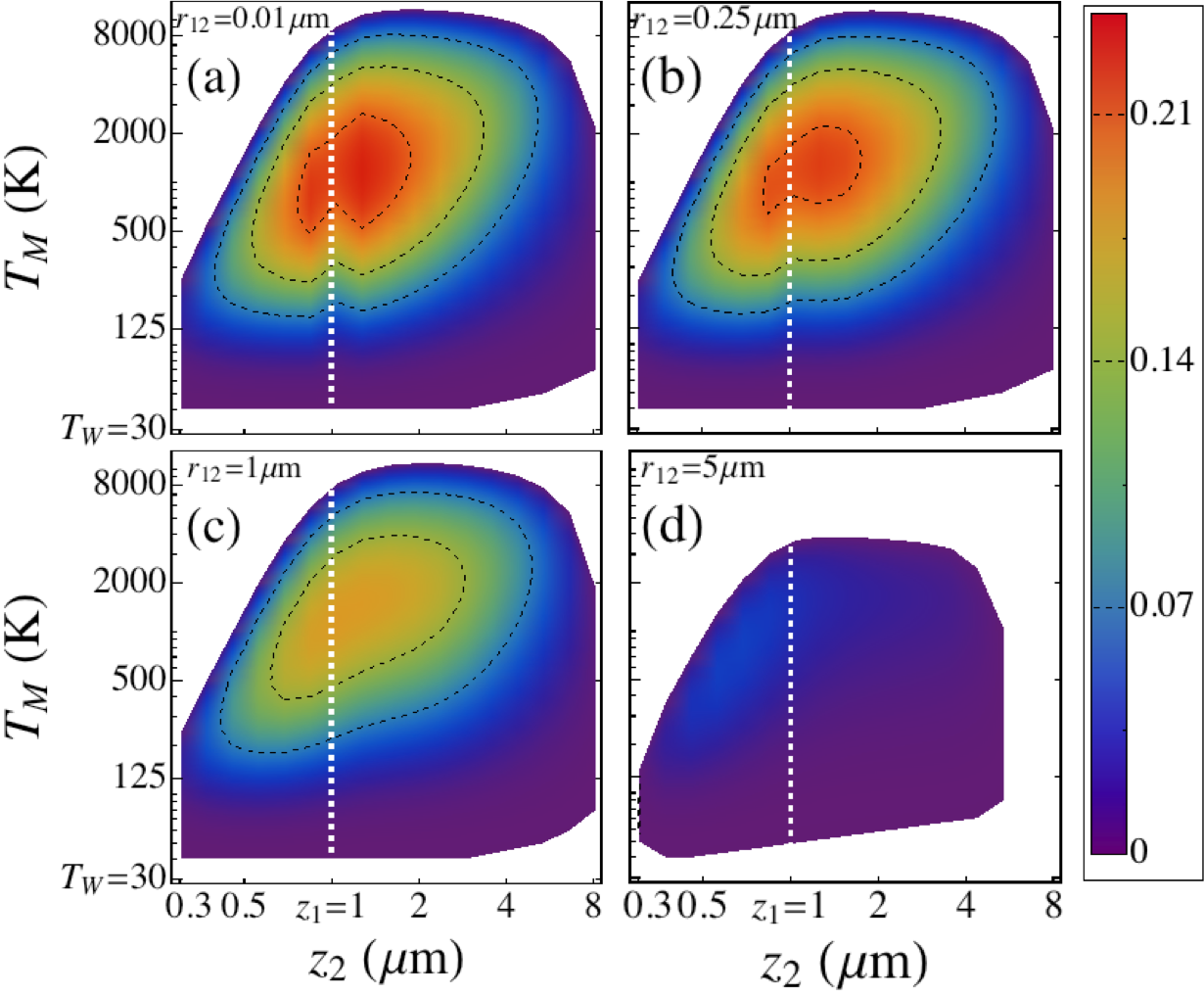} \end{center}
\caption{\label{fig:FigCvsTMvsZ2}\footnotesize Density plot of $C$ vs $z_2$ and $T_{\mathrm{M}}$, for four different values of  $r_{12}$: 0.01 $\mu$m (a), 0.25 $\mu$m (b), 1 $\mu$m (c) and 5 $\mu$m (d). The other parameters are $z_1 {\simeq 1.04} \,\mu$m, $T_{\mathrm{W}}= 30$ K, $\omega=0.3\, \omega_r$, $\delta=0.01 \mu$m. The white zones correspond to $C=0$. The two electric dipoles  are  identical and perpendicular to the slab. The white lines correspond to the case $z_2=z_1$.}
\end{figure}
 \begin{figure}[t!]
\begin{center} \includegraphics[width=0.65\textwidth]{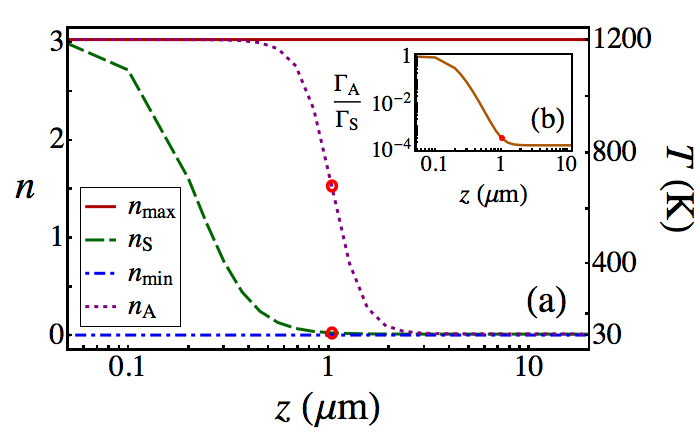} \end{center}
\caption{\label{fig:figuraNvsZ}\footnotesize   $n_{\mathrm{S}}$, $n_{\mathrm{A}}$, $n_{min}=n(0.3 \omega_r, T_{\mathrm{W}})$,  $n_{max}=n(0.3 \omega_r, T_{\mathrm{M}})$, and $\Gamma_{\mathrm{A}}/\Gamma_{\mathrm{S}}$ (inset) vs $z=z_1=z_2.$ Values of paramters: $T_{\mathrm{W}}= 30 $ K, $T_{\mathrm{M}}= 1200 $ K, $\delta=0.01 \,\mu$m and  $r_{12}$=0.25 $\mu$m. The two electric dipoles  are  identical and perpendicular to the slab. The temperatures corresponding to the values of $n$ are also indicated. }
\end{figure}

In figure \ref{fig:FigCvsTMvsZ2}  we plot the  steady concurrence as a function of the position of the second qubit $z_2$ and of the slab temperature $T_{\mathrm{M}}$ for four different values of $r_{12}$. From (a) to (d) the two-qubit distance $[r_{12}^2+(z_1-z_2)^2]^{1/2}$ increases leading to a progressive decrease of the values of concurrence generated. A maximum of $C \simeq 0.24$ is obtained in part (a) for $z_2 \simeq 1.3 \mu$m and $T_{\mathrm{M}} \simeq 1100$ K. The white lines correspond to the case $z_2=z_1$ for which equation \eref{Cote} holds for concurrence. In part (a), the maximum along the white curve is $C\sim 0.222$ in correspondence to  $\Gamma_\mathrm{A}/\Gamma_\mathrm{S} \sim 4.6\times10^{-7}$, $n_\mathrm{S}\sim 0.02$ and $n_\mathrm{A}\sim 1.56$ which correspond to effective temperatures for the two decay channels  $T_\mathrm{S}\sim 90$ K and $T_\mathrm{A}\sim 690$ K. We observe that very high  temperatures are considered in this plot only to highlight the entire region  where steady entanglement is present. At unphysical temperatures (e.g. above the melting temperature), the plot is only indicative of what would occur if a different material was chosen such that similar values of $ \varepsilon (\omega) $ [for SiC it is $\varepsilon (0.3 \,\omega_r) \sim 10.3 + 0.00721 i $] were encountered at lower frequencies. In this case, a similar behavior for steady concurrence at lower temperatures is expected. However, we remark that in our case values of $C$ higher than 0.14 are already present at $T_\mathrm{M} \simeq 500$ K in part (a).

In figure \ref{fig:figuraNvsZ}  we discuss the behavior of $n_{\mathrm{S}}$, $n_{\mathrm{A}}$ and $\Gamma_{\mathrm{A}}/\Gamma_{\mathrm{S}}$, appearing in  \eref{Cote}, as a function of $z$. We plot in part (a) $n_{\mathrm{S}}$ and $n_{\mathrm{A}}$ as a function of $z=z_1=z_2$ and compare them with the value of $n(\omega,T)$ computed at the minimal (here $T_{\mathrm{W}}= 30 $ K) and maximal (here $T_{\mathrm{M}}= 1200 $ K) temperature considered. The temperatures and the other parameters are equal to the ones giving the maximum of concurrence in figure \ref{fig:FigCvsTMvsZ2} (b) along the white line. In the inset [part (b)] we plot $\Gamma_{\mathrm{A}}/\Gamma_{\mathrm{S}}$ as a function  of $z=z_1=z_2$. The plot evidences that near $z \simeq 1 \mu$m, both conditions to reach high values of $C$ are satisfied: small values of $n_\mathrm{S}$ and  $\Gamma_{\mathrm{A}}/\Gamma_{\mathrm{S}}$ in correspondence with high enough values of $n_\mathrm{A}$ (see also figure \ref{fig:figuraconcorrenza} for a comparison).

In figure \ref{fig:FigAng2}, we analyze the dependence of steady concurrence on dipole orientations. In general, higher results are obtained when the the two dipoles are parallel. We use again the set of parameters corresponding to the maximum in figure \ref{fig:FigCvsTMvsZ2} (b) along the white line, which is obtained for dipoles along the $z$ axis. We show how concurrence decreases by changing the dipole directions towards the $x$ axis  [part (a)] always lying on the $xz$ plane and then towards the $y$ axis [part (b)]  always lying one the $xy$  [see insets in Figs. \ref{fig:FigAng2} (a)-(b)]. From part (b) it emerges that aligning the dipoles direction to the interatomic axis (which here is the $x$ axis) is the optimal choice in the  $xy$ plane inducing a lack of symmetry in this plane between the $x$ and $y$ directions.

\subsection{Dynamics}

Here we discuss the dynamical behavior of two-qubit density matrix elements and concurrence out of thermal equilibrium, also making comparisons with the thermal equilibrium case. This analysis is performed by solving numerically the evolution governed by  \eref{master equation 6}.

\begin{figure}[t!]
\begin{center} \includegraphics[width=0.98\textwidth]{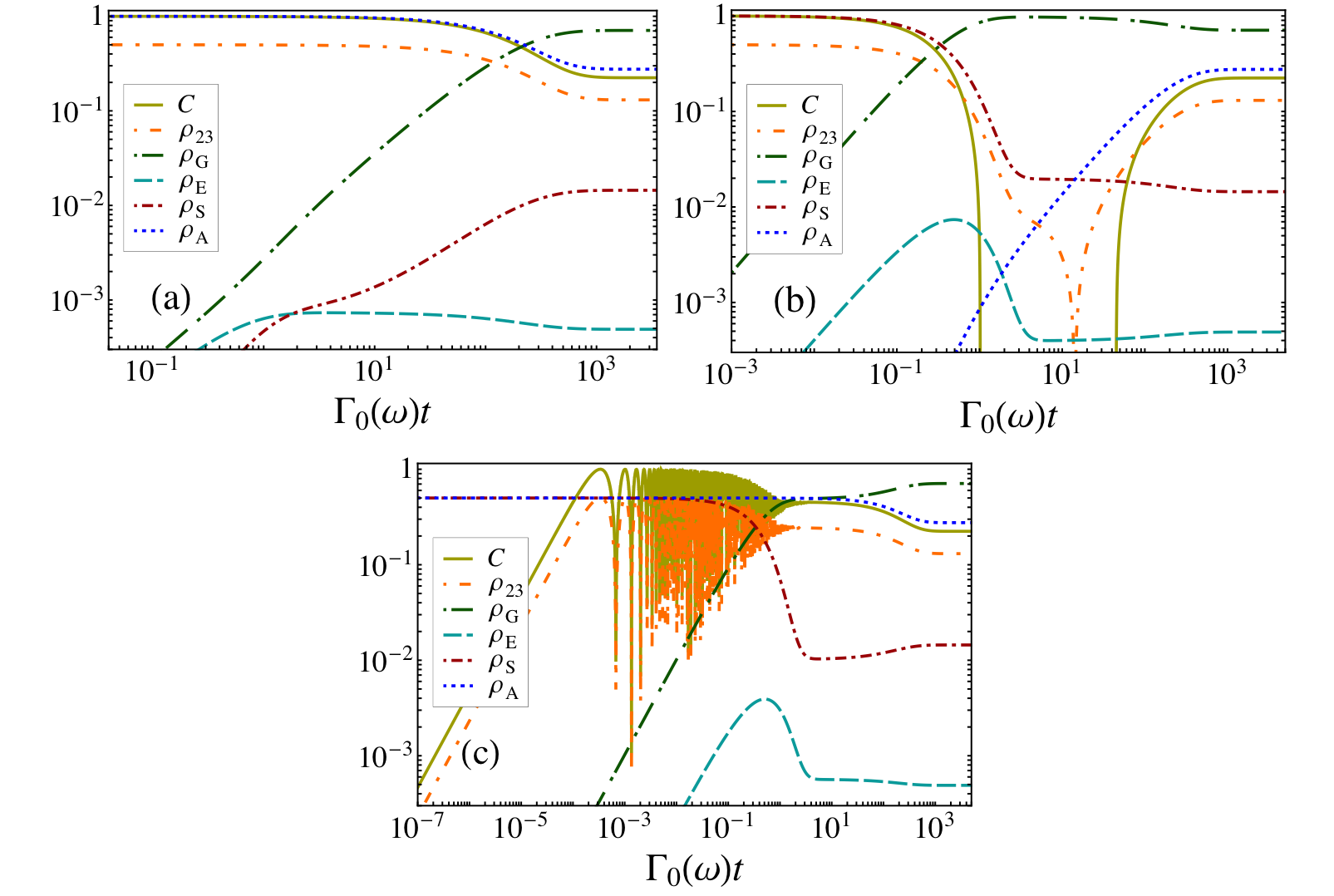}\end{center}
\caption{\label{fig:FigPopuDin}\footnotesize    $C$, $\rho_\mathrm{23}$,  $\rho_\mathrm{G}$, $\rho_\mathrm{E}$, $\rho_\mathrm{S}$ and $\rho_\mathrm{A}$  (as indicated in the legend) vs dimensionless time $\Gamma_0(\omega)t$ in the case in which the two qubits are initially prepared in the  antisymmetric state $\ket{\mathrm{A}}$ [part (a)], the symmetric state $\ket{\mathrm{S}}$ [part (b)] and the factorized state $\ket{\mathrm{2}}$ [part (c)]. The parameters are fixed as $T_{\mathrm{W}}= 30 $ K, $T_{\mathrm{M}}= 1300$ K, $\delta=0.01 \mu$m, $z_1  {\simeq 1.04}\, \mu$m, $z_2\simeq 1.28 \,\mu$m,  $r_{12}$=0.25 $\mu$m and $\omega= 0.3 \omega_r$.  The two electric dipoles  are  identical and perpendicular to the slab.}
\end{figure}

In figure \ref{fig:FigPopuDin}  we plot several density matrix elements and concurrence as a function of dimensionless time $\Gamma_0(\omega)t$. Parts (a) and (b) concern the case of maximally entangled initial states, respectively the antisymmetric state $\ket{\mathrm{A}}$ in (a) and the symmetric $\ket{\mathrm{S}}$ in (b) (see values of parameters in the caption of the figure). A quite different dynamical behavior is pointed out. While starting from  $\ket{\mathrm{A}}$ entanglement is just preserved at an high value, starting from  $\ket{\mathrm{S}}$ concurrence first decreases (going to zero) mainly because of the decrease of $\rho_{\mathrm{S}}$ and then revives because of the increase of $\rho_{\mathrm{A}}$. Dynamical creation of entanglement is yet more evident in part (c) where the initial state is the factorized state $\ket{2}$. In this case, concurrence is initially zero and increases because of the mediated interaction between qubits. Oscillations of $C$ and $\rho_{23}$ are linked to the behavior of $\rho_{\mathrm{AS}(\mathrm{SA})}$ which rapidly oscillate [see  \eref{MEsymm coher}] because of the large value of $\Lambda^{12}(\omega)$ which here is equal to  $\Lambda^{12}(\omega)/\Gamma_0(\omega)\simeq -2.3\times  {10^3}$. The oscillations are present because $\rho_{\mathrm{AS}(\mathrm{SA})}$ are initially populated [$\rho_{\mathrm{AS}(\mathrm{SA})}=1/2$] while they are not in the cases plotted in parts (a) and (b). States $\ket{\mathrm{S}}$ and $\ket{\mathrm{A}}$ are initially equally populated and become different asymptotically.

\begin{figure}[t!]
\begin{center} \includegraphics[width=0.97\textwidth]{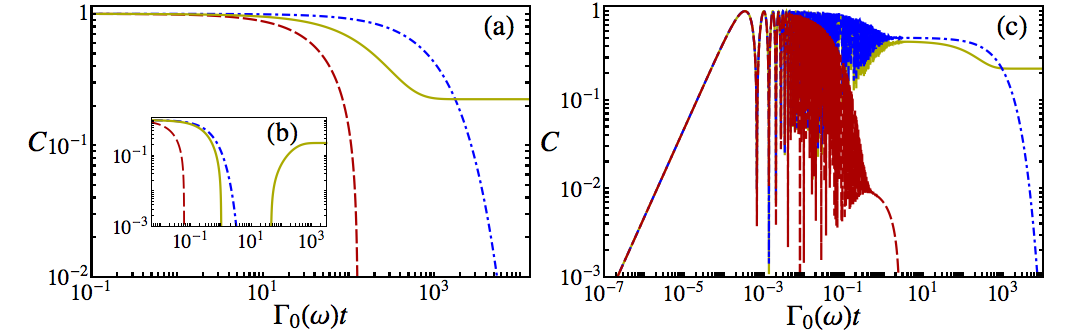}\end{center}
\caption{\label{fig:FigCdin}\footnotesize  Comparison of the dynamics of concurrence in and out of thermal equilibrium. Concurrence vs dimensionless time $\Gamma_0(\omega)t$ in the case in which the two qubits are initially prepared in the  antisymmetric state $\ket{\mathrm{A}}$ [part (a)], the symmetric state $\ket{\mathrm{S}}$ [part (b)] and the factorized state $\ket{\mathrm{2}}$ [part (c)].
The red (dashed) and blue (dotdashed) lines regard thermal equilibrium configurations at, respectively, $T_{\mathrm{max}}=T_{\mathrm{W}}=T_{\mathrm{M}}= 1300 $ K and $T_{\mathrm{min}}=T_{\mathrm{W}}=T_{\mathrm{M}}= 30 $ K while the yellow (continuous) line the out of thermal equilibrium case  $T_{\mathrm{W}}= 30 $ K and $T_{\mathrm{M}}= 1300 $ K. The other parameters are fixed as  $\delta=0.01 \mu$m, $z_1  {\simeq 1.04} \,\mu$m, $z_2 {\simeq} 1.28 \,\mu$m,  $r_{12}$=0.25 $\mu$m and $\omega= 0.3 \omega_r$.  The two electric dipoles  are  identical and perpendicular to the slab.}
\end{figure}

\begin{figure}[t!]
\begin{center} \includegraphics[width=0.98\textwidth]{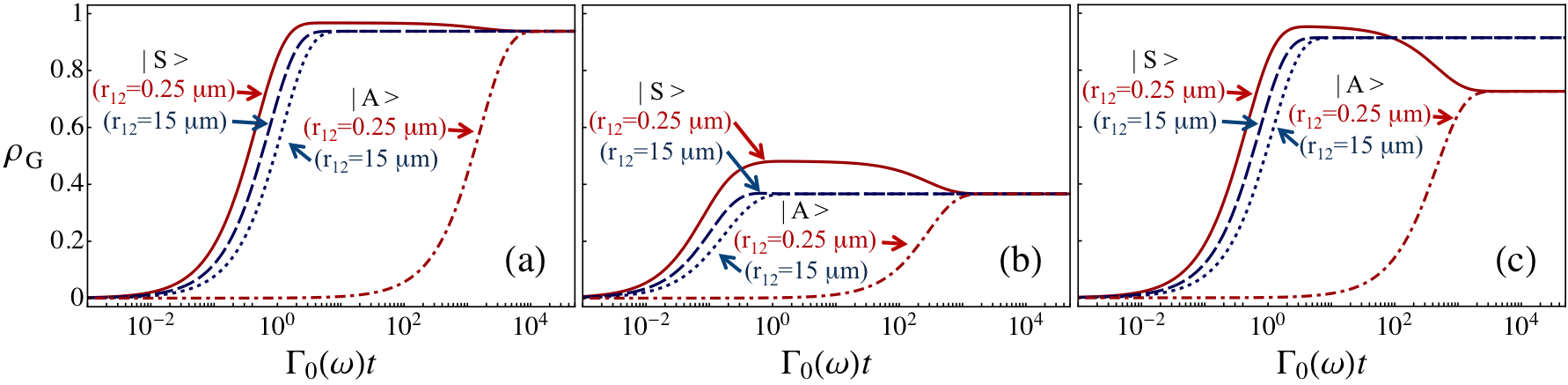}\end{center}
\caption{\label{fig:FigSubSupb}\footnotesize  Evolution of $\rho_{G}$ starting from $\ket{\mathrm{S}}$ and $\ket{\mathrm{A}}$ for two different values of $r_{12}$ [$r_{12}$=0.25 $\mu$m or $r_{12}$=15 $\mu$m]. Part (a): thermal equilibrium at $T_\mathrm{W}=T_\mathrm{M}=100$ K. Part (b): thermal equilibrium at $T_\mathrm{W}=T_\mathrm{M}=800$ K. Part (c): out of thermal equilibrium, $T_\mathrm{W}=100$ K and $T_\mathrm{M}=800$ K. Other parameters $\delta=0.01 \mu$m, $z=z_1=z_2 {\simeq 1.04}\, \mu$m and $\omega=0.3 \omega_r$. The two electric dipoles  are  identical and perpendicular to the slab. The assumptions made in Sec. \ref{subpar:analytic investigation} are thus satisfied.}
\end{figure}

In figure \ref{fig:FigCdin} we compare the evolution of concurrence out of thermal equilibrium with the evolutions at equilibrium at the minimal temperature $T_\mathrm{min}=T_\mathrm{W}=T_\mathrm{M}= 30$ K and at the  maximal temperature $T_\mathrm{max}=T_\mathrm{W}=T_\mathrm{M}=1300$ K.  Two initial maximally entangled configurations are compared, the antisymmetric state  $\ket{\mathrm{A}}$ in part (a) and the symmetric state  $\ket{\mathrm{S}}$ in part (b). At thermal equilibrium concurrence vanishes  on shorter times by increasing the  temperature, while out of equilibrium steady entanglement is present. At equilibrium, a larger decay time is observed by starting from the antisymmetric state (see also figure \ref{fig:FigSubSupb} on this subject). In part (b), out of equilibrium, concurrence decays on the same equilibrium time scale, the two-qubit state becoming separable, but it reemerges  successively. Both in (a) and (b) a large amount of the initial entanglement is thus asymptotically preserved. In part (c) the initial state is the factorized state $\ket{2}$. The main difference here is that concurrence presents strong oscillations [see comment on part (c) of figure \ref{fig:FigPopuDin}]. At thermal equilibrium entanglement eventually vanishes on a time scale similar to the one of part (a), while out of equilibrium it is maintained after its creation.

In figure \ref{fig:FigSubSupb}, we discuss the dependence of super and sub radiant effects from the presence/absence of thermal equilibrium. Here, super and sub radiance are connected to the occurrence of a decay rate larger or smaller than the one observed in the case of independent qubits, phenomenon due to the interaction of the qubits with a common environment and which depend on the nature of their initial state \cite{FicekBook2005}. In particular, we compare the evolution of the ground state population  $\rho_{\mathrm{G}}$ starting from $\ket{\mathrm{S}}$ and $\ket{\mathrm{A}}$ for two different values of $r_{12}$ (0.25 $\mu$m and 15 $\mu$m) at thermal equilibrium at $T_\mathrm{W}=T_\mathrm{M}=100$ K [part (a)] and at  $T_\mathrm{W}=T_\mathrm{M}=800$ K [part (b)] and out of thermal equilibrium for  $T_\mathrm{W}=100$ K, $T_\mathrm{M}=800$ K [part (c)]. The figure evidences super-radiant behavior when the initial state is $\ket{\mathrm{S}}$ and sub-radiant when it is $\ket{\mathrm{A}}$. The faster or slower increase of $\rho_{\mathrm{G}}$ is due to the role of $\Gamma_{12}(\pm \omega)$ which in the two channel decay rates of  \eref{channelrates}  is summed to $\Gamma (\pm \omega)$ in the  $\ket{\mathrm{S}}$ case and subtracted  in the  $\ket{\mathrm{A}}$ case. By increasing the value of $r_{12}$, $\Gamma_{12}(\pm \omega)$ decreases and the decay rates $\Gamma^{\mathrm{p(m)}}_{\mathrm{A}}$ and $\Gamma^{\mathrm{p(m)}}_{\mathrm{S}}$,
defined in the caption of figure \ref{fig:figuraschema},  tend both to the same value $\Gamma(+(-) \omega)/\Gamma_0(\omega)$, which is the decay rate in the case of single emitters. At thermal equilibrium the asymptotic state is independent on the values of $r_{12}$ while this is not the case out of thermal equilibrium, as pointed out in part (c).

We finally remark that relevant differences are expected when the Markovian and the rotating wave approximation, here adopted, are not valid. In the non-Markovian regime another source of oscillations in the dynamics of concurrence typically emerges \cite{BellomoPRL07}, while the effect of counter rotating terms is known to modify the creation  of entanglement between the two emitters \cite{Wang2013}.

\section{Conclusions \label{par:Conclusions}}

In this paper we have investigated a system made of two quantum emitters interacting with a common stationary electromagnetic field out of thermal equilibrium generated by an arbitrary body and by the surrounding walls held at fixed different temperatures. The environmental field is characterized by means of its correlation functions out of equilibrium which also depend on the scattering properties of the body.
We have derived the expressions in the absence of thermal equilibrium of the various functions governing the dissipative dynamics of the two emitters and compared them with the ones holding at thermal equilibrium. This has been done in the case of emitters characterized by an arbitrary number of levels. We have then specialized our investigation to the case of two qubits discussing the new features emerging out of thermal equilibrium.

For a restricted parameter region we have analytically shown that absence of equilibrium may lead to the generation of steady entangled states. This phenomenon has been interpreted in terms of different effective temperatures associated to two decay channels connecting the total excited and ground states via the symmetric and antisymmetric states respectively. The two-qubit dynamics can be directed towards mixed states where the antisymmetric contribution is larger than the symmetric one (or viceversa), resulting in the presence of steady entanglement. It has been found in this specific case a value of 1/3 as maximum for the concurrence, quantifying the steady entanglement.

We have then numerically investigated the general dependence of steady states and dynamics on the various parameters, without any restriction on the decay rates, in the case the body placed in proximity of the two qubits is a slab made of SiC. The dependence of steady entanglement on the two-qubit distance, their common transition frequency with respect to the slab resonances, the slab thickness, the dipoles orientations and the two involved temperatures has been discussed. Values of concurrence up to 0.24 have been found. Protection and/or generation of entanglement according to the nature of the two-qubit initial state, entangled or not, have been pointed out, also comparing entanglement dynamics in the presence or  absence of thermal equilibrium. Higher values of steady concurrence are found for transition frequencies far from the slab resonances ($\omega/\omega_r =0.3$) and small thickness ($\delta \simeq 0.01 \mu$m). Remarkably, steady entanglement can be obtained starting from configurations at thermal equilibrium and by increasing one of the two temperatures involved in the environment of the two qubits.

The possibility to observe the effects we discussed could be explored, for example, for emitters made by trapped atoms  \cite{AntezzaPRL07}  or by artificial atoms such as quantum dots or superconducting qubits, placed in proximity of a substrate held at a temperature different from that of the cell surrounding the emitters and the substrate.

\ack
Authors  thank R Messina for useful discussions and acknowledge financial support from the Julian Schwinger Foundation. MA is member of the LabEx
NUMEV.
\addcontentsline{toc}{section}{Acknowledgments}

\appendix

\section{Correlation functions} \label{app:one body}

Here we connect the  correlation functions to $T_\mathrm{M}$ and $T_\mathrm{W}$ and to the properties of the body as well.
To this purpose, we first develop the connection between \eref{Xi function} and the correlation functions in frequency space.
Using  \eref{electric field} and homogeneity in time, we have
\begin{equation}\label{Xiij}\eqalign{\fl \Xi_{ii'}^{qq'}(\omega)=&\frac{1}{\hbar^2}\int_{0}^\infty ds\int_0^{+\infty}\frac{d\omega'}{2\pi}\int_0^{+\infty}\frac{d\omega''}{2\pi}\Bigl[e^{-i(\omega''-\omega)s}\langle E_i(\mathbf{R}_q,\omega'')E_{i'}^\dag(\mathbf{R}_{q'},\omega')\rangle\\
\fl &\,+e^{i(\omega+\omega'')s}\langle E_i^\dag(\mathbf{R}_q,\omega'')E_{i'}(\mathbf{R}_{q'},\omega')\rangle\Bigr],
}\end{equation}
where we have used $\langle E_i(\mathbf{R}_q,\omega'')E_{i'}(\mathbf{R}_{q'},\omega')\rangle=\langle E_i^\dag(\mathbf{R}_q,\omega'')E_{i'}^\dag(\mathbf{R}_{q'},\omega')\rangle=0$. By using $\int_{0}^\infty ds\exp(-i\epsilon s)=\pi\delta(\epsilon) - i \mathcal{P}\frac{1}{\epsilon}$ (where $\mathcal{P}$ indicate the principal part of the integral), we obtain from previous equation and \eref{Xi function} (we assume $\omega>0$)
\begin{equation}\label{gammaij}\eqalign{
\gamma_{ii'}^{qq'}(\omega)=& \frac{1}{\hbar^2} \int_0^{+\infty}\frac{ d\omega'}{2\pi} \langle E_i(\mathbf{R}_q,\omega)E_{i'}^\dag(\mathbf{R}_{q'},\omega')\rangle, \\
 \gamma_{ii'}^{qq'}(-\omega)=& \frac{1}{\hbar^2} \int_0^{+\infty}\frac{ d\omega'}{2\pi}\langle E_i^\dag(\mathbf{R}_q,\omega)E_{i'}(\mathbf{R}_{q'},\omega')\rangle
,\\
 s_{ii'}^{qq'}(\omega)= &  \frac{1}{\hbar^2}\mathcal{P}\int_0^{+\infty}\frac{d\omega'}{2\pi}\int_0^{+\infty}\frac{d\omega''}{2\pi}
 \Bigg[\frac{\langle E_i(\mathbf{R}_q,\omega'')E_{i'}^\dag(\mathbf{R}_{q'},\omega')\rangle}{\omega-\omega''}  \\ &+\frac{\langle E_i^\dag(\mathbf{R}_q,\omega'')E_{i'}(\mathbf{R}_{q'},\omega')\rangle}{\omega+\omega''}\Bigg].
}\end{equation}
By using the decomposition in \eref{Eomega}, we obtain
\begin{equation}\label{CorrFreq}\eqalign{\fl &\langle E_i(\mathbf{R}_q,\omega)E_{i'}^\dag(\mathbf{R}_{q'},\omega')\rangle=\sum_{\phi,\phi',p,p'}\int\frac{d^2\mathbf{k}}{(2\pi)^2}\int\frac{d^2\mathbf{k}'}{(2\pi)^2}
e^{i(\mathbf{K}^\phi\cdot\mathbf{R}_q-\mathbf{K'}^{\phi'*}\cdot\mathbf{R}_{q'})}  \\& \,\times [\hat{\bbm[\epsilon]}_p^\phi(\mathbf{k},\omega)]_i[\hat{\bbm[\epsilon]}_{p'}^{\phi'}(\mathbf{k}',\omega')]_{i'}^*\langle E_p^\phi(\mathbf{k},\omega)E_{p'}^{\phi'\dag}(\mathbf{k}',\omega')\rangle ,}\end{equation}
and
\begin{equation}\label{CorrFreq-}\eqalign{&\langle E_i^\dag(\mathbf{R}_q,\omega)E_{i'}(\mathbf{R}_{q'},\omega')\rangle=\sum_{\phi,\phi',p,p'}\int\frac{d^2\mathbf{k}}{(2\pi)^2}\int\frac{d^2\mathbf{k}'}{(2\pi)^2}
 e^{-i (\mathbf{K}^{\phi*}\cdot\mathbf{R}_q-\mathbf{K'}^{\phi'}\cdot\mathbf{R}_{q'})}
 \\&\times [\hat{\bbm[\epsilon]}_p^\phi(\mathbf{k},\omega)]_i^*[\hat{\bbm[\epsilon]}_{p'}^{\phi'}(\mathbf{k}',\omega')]_{i'}\langle E_p^{\phi\dag}(\mathbf{k},\omega)E_{p'}^{\phi'}(\mathbf{k}',\omega')\rangle,}\end{equation}
where  $\omega>0$. We observe that last equation can be obtained by taking the complex conjugate of  \eref{CorrFreq} after having interchanged the operators $E_p^\phi(\mathbf{k},\omega)$ and $E_{p'}^{\phi' \dag}(\mathbf{k}',\omega')$.

We now combine equations \eref{totalfield} and \eref{sourcefields} to obtain the symmetrized correlation functions of the amplitude operator of the total field in the region of interest
\begin{equation}\label{explicitCorrB2}
\eqalign{\fl &\langle E^{+}_p(\mathbf{k},\omega)E^{+\dag}_{p'}(\mathbf{k}',\omega')\rangle_\mathrm{sym}=2\pi\delta(\omega-\omega')\frac{\omega}{2\epsilon_0c^2} \bra{p,\mathbf{k}}\Bigl[N(\omega,T_\mathrm{M})\Bigl(\mathcal{P}_{-1}^\mathrm{(pw)}-\mathcal{R}\mathcal{P}_{-1}^\mathrm{(pw)}\mathcal{R}^{\dag}\\
\fl &\,+\mathcal{R}\mathcal{P}_{-1}^\mathrm{(ew)} -\mathcal{P}_{-1}^\mathrm{(ew)}\mathcal{R}^{\dag}-\mathcal{T}\mathcal{P}_{-1}^\mathrm{(pw)}\mathcal{T}^{\dag}\Bigr) +N(\omega,T_\mathrm{W})\Bigl(\mathcal{T}\mathcal{P}_{-1}^{\mathrm{(pw)}}\mathcal{T}^{\dag}+\mathcal{R}\mathcal{P}_{-1}^{\mathrm{(pw)}}\mathcal{R}^{\dag}\Bigr)\Bigr]\ket{p',\mathbf{k}'},\\  \fl &\langle E^{+}_p(\mathbf{k},\omega)E^{-\dag}_{p'}(\mathbf{k}',\omega')\rangle_\mathrm{sym}=2\pi\delta(\omega-\omega')\frac{\omega}{2\epsilon_0c^2} N(\omega,T_\mathrm{W})\bra{p,\mathbf{k}}\mathcal{R}\mathcal{P}_{-1}^{\mathrm{(pw)}}\ket{p',\mathbf{k}'},\\ \fl
&\langle E^{-}_p(\mathbf{k},\omega)E^{+\dag}_{p'}(\mathbf{k}',\omega')\rangle_\mathrm{sym}=2\pi\delta(\omega-\omega')\frac{\omega}{2\epsilon_0c^2}N(\omega,T_\mathrm{W})\bra{p,\mathbf{k}}\mathcal{P}_{-1}^{\mathrm{(pw)}}\mathcal{R}^{\dag}\ket{p',\mathbf{k}'},\\ \fl
&\langle E^{-}_p(\mathbf{k},\omega)E^{-\dag}_{p'}(\mathbf{k}',\omega')\rangle_\mathrm{sym}=2\pi\delta(\omega-\omega')\frac{\omega}{2\epsilon_0c^2}N(\omega,T_\mathrm{W}) \bra{p,\mathbf{k}}\mathcal{P}_{-1}^{\mathrm{(pw)}}\ket{p',\mathbf{k}'}.}\end{equation}
However, in order to develop  equations \eref{CorrFreq} and \eref{CorrFreq-} we need
the non-symmetrized versions of these correlation functions. To compute them, we first remark that the source correlation functions reported in  \eref{sourcefields} have been derived using thermal-equilibrium techniques at the temperature of each source individually (see \cite{MesAntPRA11} for a detailed discussion). It follows that we can use Kubo's prescription \cite{KuboRepProgPhys66}, according to which in order to obtain $\langle AB\rangle$ from $\langle AB\rangle_\mathrm{sym}$ the replacement $N(\omega,T_i)\rightarrow\hbar\omega\bigl[1+n(\omega,T_i)\bigr]$ must be performed, whilst $\langle BA\rangle$ results from the replacement $N(\omega,T_i)\rightarrow\hbar\omega\,n(\omega,T_i)$.

 Using \eref{explicitCorrB2}  in  \eref{CorrFreq} we obtain  for the  antinormally ordered correlation functions, the form
\begin{equation} \label{totcorrfunctions}\eqalign{ \langle E_i& (\mathbf{R}_q,\omega)E_{i'}^\dag  (\mathbf{R}_{q'},\omega')\rangle
=  2\pi \delta (\omega-\omega')  \langle E_i(\mathbf{R}_q,\omega)E_{i'}^\dag(\mathbf{R}_{q'},\omega)\rangle,
}\end{equation}
being
\begin{equation}\label{funzcorr}\eqalign{
\fl &\langle E_i(\mathbf{R}_q,\omega)E_{i'}^\dag(\mathbf{R}_{q'},\omega)\rangle=\frac{\hbar\omega^2}{2\epsilon_0 c^2}\sum_{p,p'}\int\frac{d^2\mathbf{k}}{(2\pi)^2}\int\frac{d^2\mathbf{k}'}{(2\pi)^2} e^{i(\mathbf{k}\cdot\mathbf{r}_q-\mathbf{k}'\cdot\mathbf{r}_{q'})}
\\ \fl
&\,\times\bra{p,\mathbf{k}}\Bigl\{e^{i(k_z z_q-k_z^{'*}z_{q'})}
[\hat{\bbm[\epsilon]}_p^+(\mathbf{k},\omega)]_i [\hat{\bbm[\epsilon]}_{p'}^{+}(\mathbf{k}',\omega)]_{i'}^* \Bigl[\bigl[1+n(\omega,T_\mathrm{M})\bigr]
\\ \fl
&\,\times\Bigl(\mathcal{P}_{-1}^\mathrm{(pw)}-\mathcal{R}\mathcal{P}_{-1}^\mathrm{(pw)}\mathcal{R}^{\dag}  +\mathcal{R}\mathcal{P}_{-1}^\mathrm{(ew)} -\mathcal{P}_{-1}^\mathrm{(ew)}\mathcal{R}^{\dag} -\mathcal{T}\mathcal{P}_{-1}^\mathrm{(pw)}\mathcal{T}^{\dag}\Bigr)+\bigl[1+n(\omega,T_\mathrm{W})\bigr]
\\ \fl
&\,\times \Bigl(\mathcal{T}\mathcal{P}_{-1}^{\mathrm{(pw)}}\mathcal{T}^{\dag}+\mathcal{R}\mathcal{P}_{-1}^{\mathrm{(pw)}}\mathcal{R}^{\dag}\Bigr)\Bigr] +\bigl[1+n(\omega ,T_\mathrm{W})\bigr] \Bigl[e^{i(k_zz_q+k_z^{'*}z_{q'})} \\ \fl
&\,\times [\hat{\bbm[\epsilon]}_p^+(\mathbf{k},\omega)]_i[\hat{\bbm[\epsilon]}_{p'}^{-}(\mathbf{k}',\omega)]_{i'}^*\mathcal{R}\mathcal{P}_{-1}^{\mathrm{(pw)}} +e^{-i(k_zz_q+k_z^{'*}z_{q'})}[\hat{\bbm[\epsilon]}_p^-(\mathbf{k},\omega)]_i[\hat{\bbm[\epsilon]}_{p'}^{+}(\mathbf{k}',\omega)]_{i'}^*
\\ \fl
& \, \times \mathcal{P}_{-1}^{\mathrm{(pw)}}\mathcal{R}^{\dag}+e^{-i(k_z z_{q}-k_z^{'*}z_{q'})}[\hat{\bbm[\epsilon]}_p^-(\mathbf{k},\omega)]_i[\hat{\bbm[\epsilon]}_{p'}^{-}(\mathbf{k}',\omega)]_{i'}^*\mathcal{P}_{-1}^{\mathrm{(pw)}}\Bigr]\Big\}\ket{p',\mathbf{k}'}.}\end{equation}

We observe that the normally  ordered correlation functions $\langle E_i^\dag(\mathbf{R}_q,\omega)E_{i'}(\mathbf{R}_{q'},\omega')\rangle$ of \eref{CorrFreq-} are  obtained from the two previous equations by replacing $\bigl[1+n(\omega,T_i)\bigr]$ with $n(\omega,T_i)$ and by taking the complex conjugate. Equation \eref{gammaij} finally becomes
\begin{equation} \label{gammaij2} \eqalign{ \fl  \gamma_{ii'}^{qq'}(\omega)=&\frac{1}{\hbar^2} \langle E_i(\mathbf{R}_q,\omega)E_{i'}^\dag(\mathbf{R}_{q'},\omega)\rangle, \quad \gamma_{ii'}^{qq'}(-\omega)=\frac{1}{\hbar^2} \langle E_i^\dag(\mathbf{R}_q,\omega)E_{i'}(\mathbf{R}_{q'},\omega)\rangle, \\
 \fl s_{ii'}^{qq'}(\omega)=& \frac{1}{\hbar^2} \mathcal{P}\int_0^{+\infty}\frac{d\omega'}{2\pi}\Bigg[\frac{\langle E_i(\mathbf{R}_q,\omega')E_{i'}^\dag(\mathbf{R}_{q'},\omega')\rangle}{\omega-\omega'} +\frac{\langle E_i^\dag(\mathbf{R}_q,\omega')E_{i'}(\mathbf{R}_{q'},\omega')\rangle}{\omega+\omega'}\Bigg].
}\end{equation}

\section{Absence of matter} \label{app:absence of matter}

Here, we treat explicitly the case when there is no body close to the two emitters.
In this case we have in \eref {explicitCorrB2} $\mathcal{T}=1$, $\mathcal{R} =0$, or equivalently in \eref{RT1slab} $\rho_{p}(\mathbf{k},\omega)=0$ and $\tau_{p}(\mathbf{k},\omega)=1$ [$\varepsilon(\omega)=1$].  It follows that the integral matrices in \eref{integrals}
reduce to $[A^{qq'}(\omega)]_{ii'} =[B^{qq'}(\omega)]_{ii'}$  and  $ [C^{qq'}(\omega)]_{ii'}=[D^{qq'}(\omega)]_{ii'}=0$, so that $ [\alpha^{qq'}_\mathrm{M}(\omega)]_{ii'}=0$ and $ [\alpha^{qq'}_\mathrm{W}(\omega)]_{ii'}$ (we choose the $x$ axis along the inter-atomic axis so that in \eref{alphaslab} $z_q=z_{q'}=0$ and $\mathbf{R_{q}}-\mathbf{R}_{q'}=\{r_{qq'},0,0\}$):
\begin{equation}\eqalign{\label{alphawithoutmatter}
\fl [\alpha^{qq'}_\mathrm{W}(\omega)]_{ii'}= \frac{3c}{4\omega} \sum_p\int_0^{\frac{\omega}{c}}\frac{k\,dk}{k_z}  \mathrm{Re}{ [N_p^{++}(k,\omega)]_{ii'}}
= (\alpha^{qq'}_\mathrm{W})_\parallel \tilde{e}_\parallel  +   (\alpha_W)_\perp \Bigl( \tilde{e}_{\perp (y)}+\tilde{e}_{\perp (z)}\Bigr) ,
}\end{equation}
where $\tilde{e}_\parallel$, $\tilde{e}_{\perp (y)}$ and $\tilde{e}_{\perp (z)}$ are respectively unit vectors along the parallel and the perpendicular  directions to the inter-atomic axe $(x)$ and
\begin{equation}\eqalign{\label{alphawithoutmatter2}
(\alpha^{qq'}_\mathrm{W})_\parallel = 3\Bigl( \frac{ \sin \tilde{r} }{\tilde{r}^3 }-\frac{ \cos \tilde{r} }{\tilde{r}^2 }\Bigr), \quad
(\alpha^{qq'}_\mathrm{W})_\perp= \frac{3}{2}\Bigl( \frac{ \sin \tilde{r} }{\tilde{r} }+\frac{\cos \tilde{r} }{\tilde{r}^2 }-\frac{ \sin \tilde{r} }{\tilde{r}^3 } \Bigr),
}\end{equation}
where $\tilde{\mathbf{r}}=(\mathbf{R_{q}}-\mathbf{R}_{q'})\omega/c$ and  $\tilde{r}= |\tilde{\mathbf{r}}|$.
Using the two previous equations and  \eref{gammaij2} and \eref{funzcorr2}, $\Gamma^{qq'}(\omega)$ of  \eref{me parameters N} can be cast under the form, by introducing $\hat{\mathbf{r}}= \tilde{\mathbf{r}}/\tilde{r}$,
\begin{equation}\eqalign{\label{gammawithoutmatter0}
\fl \Gamma^{qq'}(\omega) = &
 \sqrt{\Gamma_0^q(\omega)\Gamma_0^{q'}(\omega) } [1+n(\omega,T_\mathrm{W})]   \Bigl\{   (\tilde{\mathbf{d}}_{mn}^{q\,*}\cdot  {\hat{\bf{r}}} ) (\tilde{\mathbf{d}}_{m'n'}^{q'}\cdot {\hat{\bf{r}}}) (\alpha^{qq'}_\mathrm{W})_\parallel\\ \fl
&\,+ \bigl[ \tilde{\mathbf{d}}_{mn}^{q\,^*} \cdot \tilde{\mathbf{d}}_{m'n'}^{q'}-
(\tilde{\mathbf{d}}_{mn}^{q\,*}\cdot {\hat{\bf{r}}} ) \;(\tilde{\mathbf{d}}_{m'n'}^{q'}\cdot  {\hat{\bf{r}}})\bigr] (\alpha^{qq'}_\mathrm{W})_\perp \Bigr\} ,
}\end{equation}
which does not depend anymore on the reference system chosen to derive  \eref{alphawithoutmatter2}.
In order to compare previous result with known expressions, let us consider the case of two qubits in vacuum with dipoles  parallel  between them  (direction $\tilde{\mathbf{d}}$) with different modulus $|\mathbf{d}_q|\neq |\mathbf{d}_{q'}|$. In this case,
previous equation reduces to the form (see for instance \cite{Agarwal1974,FicekBook2005})
\begin{equation}\eqalign{\label{gammawithoutmatter}
\fl \Gamma^{qq'}(\omega)& =
\sqrt{\Gamma_0^q(\omega)\Gamma_0^{q'}(\omega) } [1+n(\omega,T_\mathrm{W})]   \frac{3}{2} \Bigl\{  \Bigl[1-  (\mathbf{\tilde{d}}\cdot {\hat{\mathbf{r}}})^2 \Bigr]  \frac{\sin \tilde{r} }{\tilde{r} }  \\ \fl &+ \Bigl[1-  3 (\mathbf{\tilde{d}}\cdot {\hat{\mathbf{r}}})^2  \Bigr]
\Bigl( \frac{\cos \tilde{r} }{\tilde{r}^2 }-\frac{\sin \tilde{r} }{\tilde{r}^3 }\Bigr)\Bigr\}.
}\end{equation}

\section{Green's function}\label{par:Green function}

Here we connect the approaches based on field correlations functions and on Green's function in order to develop the expression for $\Lambda^{qq'}(\omega)$ of
\eref{lambda2}.
At thermal equilibrium the correlators of the total electromagnetic field outside the body follow from the fluctuation-dissipation theorem
\cite{Landau63}
\begin{equation}\label{FluDiss}\eqalign{ \fl & \langle E^\mathrm{(tot)}_i(\mathbf{R}_q,\omega)  E_{i'}^{\mathrm{(tot)}\dag}(\mathbf{R}_{q'},\omega')\rangle=2 \pi\delta(\omega-\omega') 2 \hbar [1+n(\omega,T)\Bigr] \mathrm{Im} \,G_{ii'}(\mathbf{R}_q,\mathbf{R}_{q'},\omega).}\end{equation}
In  \eref{FluDiss}  $G_{ii'}(\mathbf{R}_q,\mathbf{R}_{q'},\omega)$ is the $ii'$
component of the Green's function of the system, solution of the
differential equation (for two arbitrary points $\mathbf{R}$ and $\mathbf{R}'$)
\begin{equation}\Bigl[\nabla_\mathbf{R}\times\nabla_\mathbf{R}-\frac{\omega^2}{c^2}\varepsilon(\omega,\mathbf{R})\Bigr]\mathbb{G}(\mathbf{R},\mathbf{R}',\omega)=\frac{\omega^2}{\epsilon_0c^2}\,\mathbb{I}\,\delta(\mathbf{R}-\mathbf{R}')\end{equation}
being $\mathbb{I}$ the identity dyad and
$\varepsilon(\omega,\mathbf{R})$ the dielectric function of the
medium. The property \eref{FluDiss} does not hold in the case of
a  nonequilibrium configuration.
The comparison between  \eref{totcorrfunctions} and \eref{funzcorr2} at equilibrium $T_\mathrm{W}=T_\mathrm{M}=T$, and  \eref{FluDiss}  gives
 \begin{equation}\label{IMofG}
 \mathrm{Im}\, G_{ii'} (\mathbf{R}_q,\mathbf{R}_{q'},\omega)=\frac{\omega^3}{3\pi \epsilon_0c^3}\frac{[\alpha_\mathrm{W}^{q q'}(\omega)]_{ii'} +[\alpha_\mathrm{M}^{q q'}(\omega)]_{ii'}}{2}.
 \end{equation}

Once stated this connection, which is used in  \eref{lambda2}, we need to compute the real part of the Green's function to develop equation \eref{lambda3}.
Following appendix C of \cite{MesAntPRA11}, the $ii'$ component of the Green's function
for two arbitrary points $\mathbf{R}=\{\mathbf{r}, z\}$ and $\mathbf{R}'=\{\mathbf{r}', z'\}$ on the right side of the body
reads like
\begin{equation}\label{GS1}\eqalign{ \fl &G_{ii'}(\mathbf{R},\mathbf{R}',\omega)=G^{(0)}_{ii'}(\mathbf{R},\mathbf{R}',\omega)+G^{\mathrm{(R)}}_{ii'}(\mathbf{R},\mathbf{R}',\omega),\\
\fl &G^{(0)}_{ii'}(\mathbf{R},\mathbf{R}',\omega)=\frac{i\omega^2}{2\epsilon_0c^2}\sum_p\int\frac{d^2\mathbf{k}}{(2\pi)^2}\exp\bigl[i\mathbf{k}\cdot\bigl(\mathbf{r}-\mathbf{r}'\bigr)\bigr]\frac{1}{k_z}\Bigl[\theta(z-z')[\hat{\bbm[\epsilon]}^+_p(\mathbf{k},\omega)]_i
\\ \fl
&\times[\hat{\bbm[\epsilon]}^+_p(\mathbf{k},\omega)]_{i'} \exp\bigl[ik_z\bigl(z-z'\bigr)\bigr]+\theta(z'-z)[\hat{\bbm[\epsilon]}^-_p(\mathbf{k},\omega)]_i[\hat{\bbm[\epsilon]}^-_p(\mathbf{k},\omega)]_{i'}\exp\bigl[ik_z\bigl(z'-z\bigr)\bigr]\Bigr],\\ \fl
&G^{\mathrm{(R)}}_{ii'}(\mathbf{R},\mathbf{R}',\omega)=\frac{i\omega^2}{2\epsilon_0c^2}\sum_{pp'}\int\frac{d^2\mathbf{k}}{(2\pi)^2}\int\frac{d^2\mathbf{k}'}{(2\pi)^2} \exp\bigl[i\bigl(\mathbf{k}\cdot\mathbf{r}-\mathbf{k}'\cdot\mathbf{r}'\bigr)\bigr] \\ \fl
&\times\frac{1}{k'_z}[\hat{\bbm[\epsilon]}^+_p(\mathbf{k},\omega)]_i[\hat{\bbm[\epsilon]}^{-}_{p'}(\mathbf{k}',\omega)]_{i'} \exp\bigl[i(k_zz+k'_zz'\bigr)\bigr]\bra{\mathbf{k},p}\mathcal{R}\ket{\mathbf{k}',p'},}\end{equation}
where $\theta$ is the Heaviside step function [$\theta(x)=1$ for $x>0$
and $\theta(x)=0$ elsewhere] and  $G_{ii'}(\mathbf{R},\mathbf{R}',\omega)$ has been divided in a free term, $G^{(0)}_{ii'}(\mathbf{R},\mathbf{R}',\omega)$, independent of the scattering operators, and  a reflected one, $G^{\mathrm{(R)}}_{ii'}(\mathbf{R},\mathbf{R}',\omega)$, proportional to $\mathcal{R}$.
With regards to the imaginary part of $G_{ii'} (\mathbf{R}_q,\mathbf{R}_{q'},\omega)$ it is possible to check starting from \eref{GS1} that equation \eref{IMofG} is verified.

Concerning the real part of $G_{ii'} (\mathbf{R}_q,\mathbf{R}_{q'},\omega)$, to derive its expression
we will make use
of the properties of the polarization unit vectors,
\begin{equation}\label{PropEps}\eqalign{\fl \hat{\bbm[\epsilon]}_\TE^\phi(-\mathbf{k},\omega)&=-\hat{\bbm[\epsilon]}_\TE^\phi(\mathbf{k},\omega),\quad\hat{\bbm[\epsilon]}_\TE^{-\phi}(\mathbf{k},\omega)=\hat{\bbm[\epsilon]}_\TE^\phi(\mathbf{k},\omega), \quad
\Bigl(\hat{\bbm[\epsilon]}_\TE^\phi(\mathbf{k},\omega)\Bigr)^*=\hat{\bbm[\epsilon]}_\TE^\phi(\mathbf{k},\omega),\\ \fl \hat{\bbm[\epsilon]}_\TM^\phi(-\mathbf{k},\omega)&=\hat{\bbm[\epsilon]}_\TM^{-\phi}(\mathbf{k},\omega),\quad
\Bigl(\hat{\bbm[\epsilon]}_\TM^\phi(\mathbf{k},\omega)\Bigr)^*=\cases{\hat{\bbm[\epsilon]}_\TM^\phi(\mathbf{k},\omega)
 &$ k_z\in\mathbb{R}$ \\
\hat{\bbm[\epsilon]}_\TM^{-\phi}(\mathbf{k},\omega) &$k_z \notin \mathbb{R}$\\},}\end{equation}
and of the  reciprocity
relations of scattering operators presented in appendix D of \cite{MesAntPRA11}
\begin{equation}\label{reciprocity}\frac{1}{k^{'*}_z}(-1)^{p+p'}\bra{-\mathbf{k}',p'}\mathcal{R}^{\phi\dag}\ket{-\mathbf{k},p}=\frac{1}{k^*_z}\bra{\mathbf{k},p}\mathcal{R}^{\phi\dag}\ket{\mathbf{k}',p'}.\end{equation}
Starting from the free term $G^{(0)}_{ii'}$  in \eref{GS1}, its real part can be written as the sum of two terms coming, respectively, from the propagative and evanescent sector
(a change of variable from $\mathbf{k}$ to $-\mathbf{k}$ is done in the terms obtained by complex conjugation, we make use
of \eref{PropEps} and we choose the interatomic axis along the $z$ direction):
\begin{equation}\label{ReG02}\eqalign{\fl
&\mathrm{Re}\, G^{(0)}_{ii'}(\mathbf{R},\mathbf{R}',\omega)_{PW}=\frac{i \omega^2}{4\epsilon_0c^2}  \frac{1}{(2\pi)^2}\int_{0}^{\frac{\omega}{c}} \frac{k d k}{k_z} [R(k,\omega)]_{ii'}
\Bigl\{  \theta(z-z') \Bigl[\exp\bigl[i k_z \bigl(z-z'\bigr)]  \\  \fl &\,  -\exp\bigl[-i k_z \bigl(z-z'\bigr)] \Bigr] +\theta(z'-z) \Bigl[ \exp\bigl[i k_z \bigl(z'-z\bigr)\bigr]-\exp\bigl[-i k_z \bigl(z'-z\bigr)\bigr]\Bigr]\Bigr\} ,\\ \fl
&\mathrm{Re}\, G^{(0)}_{ii'}(\mathbf{R},\mathbf{R}',\omega)_{EW}=\frac{i \omega^2}{4\epsilon_0c^2}  \frac{2}{(2\pi)^2} \! \int_{\frac{\omega}{c}}^{+\infty}\! \frac{k d k}{k_z} [R(k,\omega)]_{ii'} \\  \fl &\,\times\Bigl\{\theta(z-z')\exp\bigl[i k_z \bigl(z-z'\bigr)]+\theta(z'-z)\exp\bigl[i k_z \bigl(z'-z\bigr)\bigr] \Bigr\} ,
}\end{equation}
where we have used the angular integrals
\begin{equation}\label{angularintegrals0}\eqalign{
[R(k,\omega)]_{ii'}&=\sum_p\int_0^{2\pi}d\theta [\hat{\bbm[\epsilon]}^{+(-)}_p(\mathbf{k},\omega)]_i[\hat{\bbm[\epsilon]}^{+(-)}_{p}(\mathbf{k},\omega)]_{i'} ,
}\end{equation}
being the matrix $R(k,\omega)$ diagonal with $ [R(k,\omega)]_{11} =[R(k,\omega)]_{22}= \pi( 2-c^2 k^2/\omega ^2)$ and  $[R(k,\omega)]_{33} =2 \pi c^2 k^2/\omega^2$.
The integral in $\mathrm{Re}\, G^{(0)}_{ii'}(\mathbf{R},\mathbf{R}',\omega)_{PW}$ gives two terms, one erasing exactly the integral in $\mathrm{Re} \,G^{(0)}_{ii'}(\mathbf{R},\mathbf{R}',\omega)_{EW}$, and the second being equal to (we distinguish diagonal elements perpendicular and parallel to the interatomic axis)
\begin{equation}\label{ReG03}\eqalign{
&\mathrm{Re}\, G^{(0)}_{\perp}(\mathbf{R},\mathbf{R}',\omega)=\frac{1}{4 \pi}\frac{ \omega^3}{\epsilon_0c^3}
 \Bigl[\frac{(\tilde{r}^2-1) \cos \tilde{r} -  \tilde{r} \sin \tilde{r}}{\tilde{r}^3} \Bigr] \\
 &\mathrm{Re} \,G^{(0)}_{\parallel}(\mathbf{R},\mathbf{R}',\omega)=\frac{1}{2 \pi}\frac{ \omega^3}{\epsilon_0c^3}
 \Bigl[\frac{ \cos \tilde{r} +  \tilde{r} \sin \tilde{r}}{\tilde{r}^3}\Bigr],
}\end{equation}
where $\tilde{r}= |\mathbf{R}-\mathbf{R}'|\omega/c$ and we named $\mathrm{Re} \,G^{(0)}_{xx}=\mathrm{Re} \,G^{(0)}_{yy}=\mathrm{Re}\, G^{(0)}_{\perp}$ and  $\mathrm{Re} \,G^{(0)}_{zz}=\mathrm{Re} \,G^{(0)}_{\parallel}$.

Function $\Lambda^{qq'}(\omega)$ of  \eref{lambda3}  can be thus decomposed in two parts, $\Lambda^{qq'}(\omega)=\Lambda_0^{qq'}(\omega)+\Lambda_R^{qq'}(\omega)$, connected to $G^{(0)}(\mathbf{R}_q,\mathbf{R}_{q'},\omega)$ and $G^{\mathrm{(R)}}(\mathbf{R}_q,\mathbf{R}_{q'},\omega)$,  $ \Lambda_0^{qq'}(\omega)$ being
\begin{equation}\label{freelambageneral}\eqalign{ \fl
\Lambda_0^{qq'}(\omega)=&
-\sqrt{\Gamma_0^q(\omega)\Gamma_0^{q'}(\omega)} \frac{3}{4} \Bigl\{ \bigl[ \tilde{ \mathbf{d}}_{mn}^{q\,*} \cdot \tilde{\mathbf{d}}_{m'n'}^{q'} -
(\tilde{\mathbf{d}}_{mn}^{q\,*}\cdot  {\hat{\bf{r}}} ) \;(\tilde{\mathbf{d}}_{m'n'}^{q'}\cdot  {\hat{\bf{r}}})\bigr] \\ \fl&\, \times
 \Bigl[\frac{(\tilde{r}^2-1) \cos \tilde{r} -  \tilde{r} \sin \tilde{r}}{\tilde{r}^3} \Bigr]  +2 (\tilde{\mathbf{d}}_{mn}^{q\,^*}\cdot {\hat{\bf{r}}} ) \;(\tilde{\mathbf{d}}_{m'n'}^{q'}\cdot {\hat{\bf{r}}})   \Bigl[\frac{ \cos \tilde{r} +  \tilde{r} \sin \tilde{r}}{\tilde{r}^3}\Bigr] \Bigr\}, }
 \end{equation}
which has been put under a form which does not depend anymore on the reference system chosen to derive equation \eref{ReG03}.
In the case of two qubits in a vacuum  field in absence of matter with electric dipoles parallels  between them  (direction $\tilde{\mathbf{d}}$) with  $|\mathbf{d}_q|\neq |\mathbf{d}_{q'}|$, $\Lambda_0^{qq'}(\omega)$ of  \eref{freelambageneral} reduces to the known form  \cite{Agarwal1974,FicekBook2005}
\begin{equation}\label{freeLambda}
\eqalign{\fl
&\Lambda_0^{qq'}(\omega)=
\sqrt{\Gamma_0^q(\omega)\Gamma_0^{q'}(\omega) }    \frac{3}{4} \Bigl\{  \Bigl[1-  3 (\mathbf{\tilde{d}}\cdot {\hat{\mathbf{r}}})^2  \Bigr]
\Bigl( \frac{\sin \tilde{r} }{\tilde{r}^2 }+\frac{\cos \tilde{r} }{\tilde{r}^3 }\Bigr) -\Bigl[1-  (\mathbf{\tilde{d}}\cdot  {\hat{\mathbf{r}}} )^2 \Bigr]  \frac{\cos \tilde{r} }{\tilde{r} } \Bigr\},
 }\\
 \end{equation}
where we used $\mathbf{\tilde{d}}\cdot {\hat{\mathbf{r}}}=\tilde{d}_z$ and $\tilde{d}_x^2+\tilde{d}_y^2=1-(\tilde{d}_z)^2$.

We now consider the remaining part of the Green's function in \eref{GS1}.
By making  a change of variable in the terms obtained by complex conjugation,
$(\mathbf{k},\mathbf{k}')\rightarrow(-\mathbf{k},-\mathbf{k}')$, and  using the reciprocity
relations of scattering operators in \eref{reciprocity} and the properties of the polarization unit vectors \eref{PropEps}, one can obtain
\begin{equation}\label{ReGR}\eqalign{ \fl
&\mathrm{Re}\, G^{(R)}_{ii'}(\mathbf{R},\mathbf{R}',\omega) =\frac{i \omega^2}{4\epsilon_0c^2}\sum_{p,p'}\int\frac{d^2\mathbf{k}}{(2\pi)^2}\int\frac{d^2\mathbf{k}'}{(2\pi)^2}e^{i(\mathbf{k}\cdot\mathbf{r}-\mathbf{k}'\cdot\mathbf{r}')} \bra{p,\mathbf{k}}\Bigl\{e^{i(k_zz+k_z^{'*}z')}\\ \fl
&\,\times  [\hat{\bbm[\epsilon]}_p^+(\mathbf{k},\omega)]_i[\hat{\bbm[\epsilon]}_{p'}^{-}(\mathbf{k}',\omega)]_{i'}^* \mathcal{R}\mathcal{P}_{-1}^{\mathrm{(pw)}} -e^{-i(k_zz+k_z^{'*}z')}[\hat{\bbm[\epsilon]}_p^-(\mathbf{k},\omega)]_i[\hat{\bbm[\epsilon]}_{p'}^{+}(\mathbf{k}',\omega)]_{i'}^*\mathcal{P}_{-1}^{\mathrm{(pw)}}\mathcal{R}^{\dag}\\ \fl
&\, + e^{i(k_z z-k_z^{'*}z')}[\hat{\bbm[\epsilon]}_p^+(\mathbf{k},\omega)]_i[\hat{\bbm[\epsilon]}_{p'}^{+}(\mathbf{k}',\omega)]_{i'}^* \Bigl(\mathcal{R}\mathcal{P}_{-1}^\mathrm{(ew)}+\mathcal{P}_{-1}^\mathrm{(ew)}\mathcal{R}^{\dag}\Bigr)\Big\}\ket{p',\mathbf{k}'}.}\end{equation}

\end{document}